\documentclass{aa}
\usepackage[varg]{txfonts}
\usepackage{natbib}
\usepackage{graphicx}
\usepackage{times}
\bibpunct{(}{)}{;}{a}{}{,}
\newcommand{\hd}{HD~81357 }
\newcommand{\hn}{HD~81357}

\newcommand{\Mnom}{\hbox{$\mathcal{M}^{\mathrm N}_\odot$}}
\newcommand{\Rnom}{\hbox{$\mathcal{R}^{\mathrm N}_\odot$}}

\newcommand{\spefo}{{\tt SPEFO} }
\newcommand{\spefoe}{{\tt SPEFO}}
\newcommand{\phoebe}{{\tt PHOEBE} }
\newcommand{\phoebee}{{\tt PHOEBE}}
\newcommand{\fotel}{{\tt FOTEL} }

\newcommand{\korel}{{\tt KOREL} }
\newcommand{\korele}{{\tt KOREL}}

\newcommand{\pyt}{{\tt PYTERPOL} }
\newcommand{\pyte}{{\tt PYTERPOL}}

\newcommand{\tria}{\hbox{$\bigtriangleup$}}
\newcommand{\ubv}{\hbox{$U\!B{}V$}}
\newcommand{\ubvr}{\hbox{$U\!B{}V\!R$}}
\newcommand{\bv}{\hbox{$B\!-\!V$}}
\newcommand{\ub}{\hbox{$U\!-\!B$}}
\newcommand{\vr}{\hbox{$V\!-\!R$}}

\newcommand{\hp}{\hbox{$H_{\rm p}$}}

\newcommand{\p}{$\pm$}

\newcommand{\m}{$^{\rm m}\!\!.$}
\newcommand{\ANG}{\accent'27A}

\newcommand{\kms}{km~s$^{-1}$ }
\newcommand{\ks}{km~s$^{-1}$}
\newcommand{\vsin}{$v$~sin~$i$ }

\newcommand{\tef}{$T_{\rm eff}$ }
\newcommand{\teff}{$T_{\rm eff}$}
\newcommand{\lgg}{{\rm log}~$g$ }

\newcommand{\rs}{R$_{\odot}$}

\newcommand{\ha}{H$\alpha$ }

\newcommand{\hg}{H$\gamma$ }
\newcommand{\hae}{H$\alpha$}

\newcommand{\Ame}{\ANG~mm$^{-1}$}

\usepackage{color}

\begin{document}

   \title{Properties and nature of Be stars
\thanks{Based on new spectroscopic and photometric observations
from the following instruments:
CCD coud\'e spectrograph of the 2.0~m reflector of the Astronomical Institute
AS~\v{C}R, Ond\v{r}ejov, Czech Republic; CCD coud\'e spectrograph of the
1.22~m reflector of the Dominion Astrophysical Observatory, Victoria, Canada;
photoelectric photometer of the 0.65~m Cassegrain reflector of the Hvar
Observatory, Croatia, the \hp\ photometry from the ESA Hipparcos mission,
and the ASAS-SN all-sky survey $V$ photometry.}
}
\subtitle{31. The binary nature, light variability, physical elements, and
emission-line changes of HD~81357}
\author {P.~Koubsk\'y\inst{1}\and
P.~Harmanec\inst{2}\and
M.~Bro\v{z}\inst{2}\and
L.~Kotkov\'a\inst{1}\and
S.~Yang\inst{3}\and
H.~Bo\v{z}i\'c\inst{4}\and
D.~Sudar\inst{4}\and
Y.~Fr\'emat\inst{5}\and
D.~Kor\v{c}\'akov\'a\inst{2}\and
V.~Votruba\inst{6}\and
P.~\v{S}koda\inst{1}\and
M.~\v{S}lechta\inst{1}\and
D.~Ru\v{z}djak\inst{4}
}

   \offprints{P.~Koubsk\'y \,\\
               \email pavel.koubsky@asu.cas.cz}

  \institute{
  Astronomical Institute, Academy of Sciences of the Czech Republic, \hfill\break
  Fri\v{c}ova~298, CZ-251 65 Ond\v{r}ejov, Czech Republic
  \and
   Astronomical Institute of Charles University,
   Faculty of Mathematics and Physics,\hfill\break
   V~Hole\v{s}ovi\v{c}k\'ach~2, CZ-180~00 Praha~8 - Troja, Czech Republic
 \and
   Physics \& Astronomy Department, University of Victoria,\hfill\break
   PO~Box~3055~STN~CSC, Victoria,~BC, V8W~3P6, Canada
 \and
  Hvar Observatory, Faculty of Geodesy, University of Zagreb,\hfill\break
  Ka\v{c}i\'ceva~26, 10000~Zagreb, Croatia
  \and
   Royal Observatory of Belgium, Ringlaan 3, B-1180 Brussel, Belgium
  \and
  Department of Theoretical Physics and Astrophysics, Masaryk University,\hfill\break
   Kotl\'{a}\v{r}sk\'a~2,  CZ-611~37 Brno, Czech Republic
}
\date{Received \today}

  \abstract{
Reliable determination of the basic physical properties of hot emission-line
binaries with Roche-lobe filling secondaries is important for developing
the theory of mass exchange in binaries.
It is not easy, however, due to the presence of circumstellar matter.
Here, we report the first detailed investigation
of a new representative of this class of binaries, HD~81357,
based on the analysis of spectra and photometry from several observatories.
HD~81357 was found to be a double-lined spectroscopic binary and an ellipsoidal
variable seen under an intermediate orbital inclination of
$\sim(63\pm5)^\circ$, having an orbital period of 33\fd77445(41) and a~circular
orbit. From an automated comparison of the observed and synthetic spectra,
we estimate the component's effective temperatures to be 12930(540)~K
and 4260(24)~K. The combined light-curve and orbital solutions, also
constrained by a very accurate Gaia Data Release~2 parallax, give
the following values of the basic physical properties:
masses $3.36\pm0.15$ and $0.34\pm0.04$~\Mnom,
radii $3.9\pm0.2$ and $13.97\pm0.05$~\Rnom,
and a~mass ratio $10.0\pm0.5$.
Evolutionary modelling of the system including the phase of mass transfer
between the components indicated that HD~81357 is a~system observed
in the final slow phase of the mass exchange after the mass-ratio reversal.
Contrary to what has been seen for similar binaries like AU~Mon, no cyclic
light variations were found on a~time scale an~order of magnitude longer
than the orbital period.
}

\keywords{Stars: close -- Stars: binaries: spectroscopic --
          Stars: emission-line, Be --
          Stars: fundamental parameters --
          Stars: individual: HD 81357}

\authorrunning{}
\titlerunning{}
\maketitle

\section {Introduction}

\begin{table*}
\begin{center}
\caption[]{Journal of radial-velocity (RV) data sets.}\label{jourv}.
\begin{tabular}{ccccccccll}
\hline\hline\noalign{\smallskip}
 File/   &Time interval&No.   &Spectral &Spectral&S/N range&Median&Exposure\\
$\gamma$ No. &        &of    & range   &2-pixel &         &  S/N & times  \\
        &(RJD)        &RVs   & (\AA)   &res.    &         &      & (s)    \\
\noalign{\smallskip}\hline\noalign{\smallskip}
A/1&55621.58--56433.38& 26&6253--6764&12700& 57--201&109&1086--4899\\
B/1&56642.61--57118.54& 14&6262--6734&12700& 91--158&121&1761--5800\\
C/2&55621.62--57118.48& 31&8400--8900&17000& 33--84 & 62&1200--5000\\
D/3&56765.36--56853.37& 18&4273--4507&17050& 22--62 & 40&4292--15940\\
E/4&55680.76--56819.73& 27&6150--6760&21700& 16--131& 88&1415--4500\\
 \noalign{\smallskip}\hline\noalign{\smallskip}
 \end{tabular}
 \tablefoot{{\sl Sub-column $\gamma$ No.} identifies
individual systemic velocities considered in trial orbital solutions.\\
{\sl Sub-column ``File":} \ \
A: OND 2.0 m reflector, coud\'e grating spg., CCD SITe5 2030 x 800
 pixel detector, red spectra;
B: OND 2.0 m reflector, coud\'e grating spg., CCD Pylon Excelon
 2048 x 512 pixel detector, red spectra;
C: OND 2.0 m reflector, coud\'e grating spg., CCD Pylon Excelon
 2048 x 512 pixel detector, infrared spectra;
D: OND 2.0 m reflector, coud\'e grating spg., CCD Pylon Excelon
 2048 x 512 pixel detector, blue spectra;
E: DAO 1.22 m reflector, coud\'e grating McKellar spg., CCD Site4 detector,
red spectra.
{\sl Columns with the abbrevation S/N} provide the signal-to-noise ratios
of the respective spectra.\\
 }
 \end{center}
 \end{table*}

HD 81357 (BD+58$^\circ$1190, MWC~859, HIP~46377, 2MASS J09272389+5808342) is
a little studied 7\m9 star classified B8 in the HD catalogue. \citet{mwc3}
included HD~81357 in the third edition of the Mount Wilson Catalogue of B and
A emission stars and noted that on one-prism spectrograms of HD~81357 taken
in January and December, 1948, \ha was a wide, bright line, possibly double,
and superposed on a broad absorption. \citet{allen1973} obtained near-infrared
(IR) photometry of emission-line stars. For \hn, he gives only H and K magnitudes
and classifies it as ``category~X", meaning a star for which the
observed continuum distribution corresponds to that expected for its spectral
type within the observational uncertainty of the observed magnitudes and
the deduced interstellar reddening. \citet{andfer82} included HD~81357 in
their catalogue of Be stars observed with a~512-pixel television equipment
Multiphot\footnote{see \citet{adria78} for its detailed description}
attached to the Echelec spectrograph of the Haute Provence Observatory 1.52-m
reflector. A~spectrum taken on February 13, 1981 showed double-peak
\ha\ emission with $I_V=1.88$, and $I_R=1.65$.
\citet{buscombe1984} confirmed that HD~81357 had a spectral type B8e.
\citet{halbedel1996} determined its rotational velocity of 150~\kms and
spectral type B9e. She mentioned \hd under the section entitled ``Binaries"
and remarked that ``\hd also shows a curious spectrum.
It seems to be composite with an F/G secondary, though it is possible
that it has a very rich assortment of shell lines."
To encourage further study of this object, she added that
``it does seem to undergo minor velocity changes."
\cite{koeneyer2002} extracted new candidate periodic variables from the epoch
photometry of the Hipparcos catalogue. The analysis of 129 measurements
for HD~81357 yielded a~frequency of 0.03884~d$^{-1}$ ($P=25\fd747$) with
an amplitude of 0\m0189, the type of the variability being denoted as U,
meaning unsolved. \citet{wog2010} listed \hd as an Herbig Ae/Be star binary
without giving any reference. We were unable to find any report of the
presence of a strong IR excess due to dust (a criterion to distinguish
the classical and Herbig Be stars) in the bibliography of \hn, and note that
the star is located outside the zone of the star formation, which is
another defining characteristics of the Herbig Be/Ae stars.
\citet{koubskytatry}, motivated by the note of \citet{halbedel1996} and
by the results of \cite{koeneyer2002}, analysed ten spectra of HD~81357
taken in \ha and near-IR regions and measured the radial velocities (RVs)
of a number of metallic lines, clearly belonging to a late spectral
type (mainly Fe~I and Ca~II lines). They found that both - the RVs and
Hipparcos photometry - could be folded with a period of 33.8 days, thus
confirming the binary nature of the object. The corresponding semiamplitude
of RV curve was 79~\ks. No lines corresponding to a B spectral type
were reliably detected, with the exception of the \ion{H}{i} lines, filled
by emission. Here we present the first detailed study of the system,
based primarily on the new spectral and photometric observations
accumulated since Spring 2011 at four observatories.

\section{Observations and reductions}
Throughout this paper, we specify all times of observations using
reduced heliocentric Julian dates

\smallskip
\centerline{RJD = HJD -2400000.0\,.}
\smallskip
\subsection{Spectroscopy}
The star was observed at the Ond\v{r}ejov (OND),
and Dominion Astrophysical (DAO) observatories. The majority of the spectra
cover the red spectral region around \hae, but we also obtained some spectra
in the blue region around \hg and infrared spectra in the region also used
for the Gaia satellite. The journal of spectral observations is in
Table~\ref{jourv}.

The methods of spectra reductions and measurements were basically the same
as in the previous paper~30 of this series devoted to BR~CMi \citep{zarfin30}.
See also Appendix~\ref{apa} for details.

\subsection{Photometry}
The star was observed at Hvar over several
observing seasons, first in the \ubv, and later in the \ubvr\ photometric
system relative to HD~82861.
The check star HD~81772 was observed as frequently as the variable. All data
were corrected for differential extinction and carefully transformed
to the standard systems via non-linear transformation formulae.
We also used the Hipparcos \hp\ observations, transformed to the Johnson $V$
magnitude after \citet{hpvb} and recent ASAS-SN Johnson $V$ survey observations
\citep{asas2014,asas2017}. Journal of all observations is in
Table~\ref{jouphot} and the individual Hvar observations are tabulated in
Appendix~\ref{apb}, where details on individual data sets are also given.

\begin{table}
\caption[]{Journal of available photometry.}\label{jouphot}
\begin{center}
\begin{tabular}{rcrccl}
\hline\hline\noalign{\smallskip}
Station&Time interval& No. of &Passbands&Ref.\\
       &(RJD)&obs.  & \\
\noalign{\smallskip}\hline\noalign{\smallskip}
61&47879.03--48974.17& 128&$H_{\rm p}$  & 1 \\
01&55879.62--56086.36&  36&\ubv         & 2\\
01&56747.35--57116.36&  57&\ubvr        & 2\\
93&56003.85--58452.08& 209&$V$          & 3\\
\noalign{\smallskip}\hline
\end{tabular}\\
\tablefoot{{\sl Column ``Station":} The running numbers of
individual observing stations they have in the Prague / Zagreb photometric
archives:\\
01~\dots~Hvar 0.65-m, Cassegrain reflector, EMI9789QB tube;\\
61~\dots~Hipparcos all-sky $H_{\rm p}$ photometry transformed to Johnson $V$.\\
93~\dots~ASAS-SN all-sky Johnson $V$ photometry.\\
{\sl Column ``Ref."} gives the source of data:\\
1~\dots~\cite{esa97};
2~\dots~this paper;
3~\dots~\citet{asas2014,asas2017}.
}
\end{center}
\end{table}

\section{Iterative approach to data analysis}

  The binary systems in phase of mass transfer between the
components usually display rather complex spectra; besides the lines of
both binary components there are also some spectral lines related
to the circumstellar matter around the mass-gaining star
\citep[see e.g.][]{desmet2010}.
Consequently, a straightforward analysis of the spectra based on one
specific tool (e.g. 2-D cross-correlation or spectra disentangling) to obtain
RVs and orbital elements cannot be applied. One has to combine several various
techniques and data sets to obtain the most consistent solution. This is what
we tried, guided by our experience from the previous paper of this series
\citep{zarfin30}. We analysed the spectroscopic and photometric observations
iteratively in the following steps, which we then discuss in the
subsections below.

\begin{enumerate}
\item We derived and analysed RVs in several different ways to verify and
demonstrate that \hd is indeed a double-lined spectroscopic binary and
ellipsoidal variable as first reported by \citet{koubskytatry}.
\item We disentangled the spectra of both binary components in a blue spectral
region free of emission lines with the program \korel
\citep{korel1,korel2,korel3} to find out that -- like in the case of BR~CMi
\citep{zarfin30} -- a rather wide range of mass ratios gives comparably
good fits.
\item Using the disentangled component spectra as templates,
we derived 2-D cross-correlated RVs of both components with the help of
the program {\tt asTODCOR} \citep[see][for the details]{desmet2010, zarfin30}
for the best and two other plausible values of the mass ratio.
We verified that the {\tt asTODCOR} RVs of the hotter component
define a RV curve, which is in an~almost exact anti-phase to that based
on the lines of the cooler star. A disappointing finding was that the
resulting RVs of the hot component differ for the three different
\korel templates derived for the plausible range of mass ratios.
\item We used the \pyt program kindly provided by J.~Nemravov\'a
to estimate the radiative properties of both binary components from
the blue spectra, which are the least affected by circumstellar emission.
The function of the program is described in detail in \citet{jn2016}.
\footnote{The program \pyt is available with
a~tutorial at \\ https://github.com/chrysante87/pyterpol/wiki\,.}
\item Keeping the effective temperatures obtained with \pyt fixed,
and using all RVs for the cooler star from all four types of spectra together
with all photometric observations we started to search for a~plausible
combined light-curve and orbital solution. To this end we used the latest
version of the program \phoebe 1.0 \citep{prsa2005,prsa2006} and tried
to constrain the solution also by the accurate Gaia parallax of the binary.
\item As an additional check, we also used the program {\tt BINSYN}
\citep{linnell1994} to an independent verification of our results.
\end{enumerate}

\subsection{The overview of available spectra}
\begin{figure}
\centering
\includegraphics[width=9.0cm]{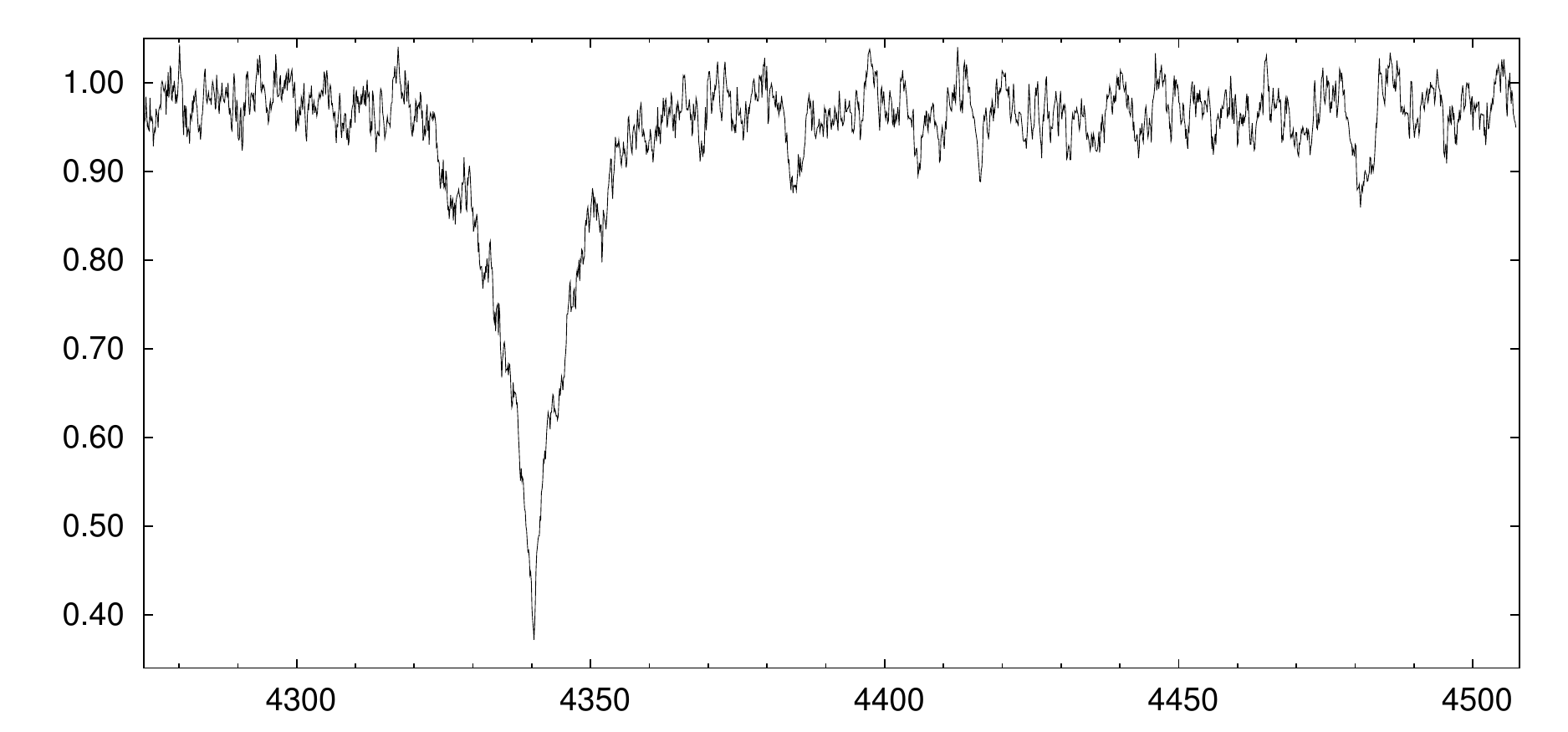}
\includegraphics[width=9.0cm]{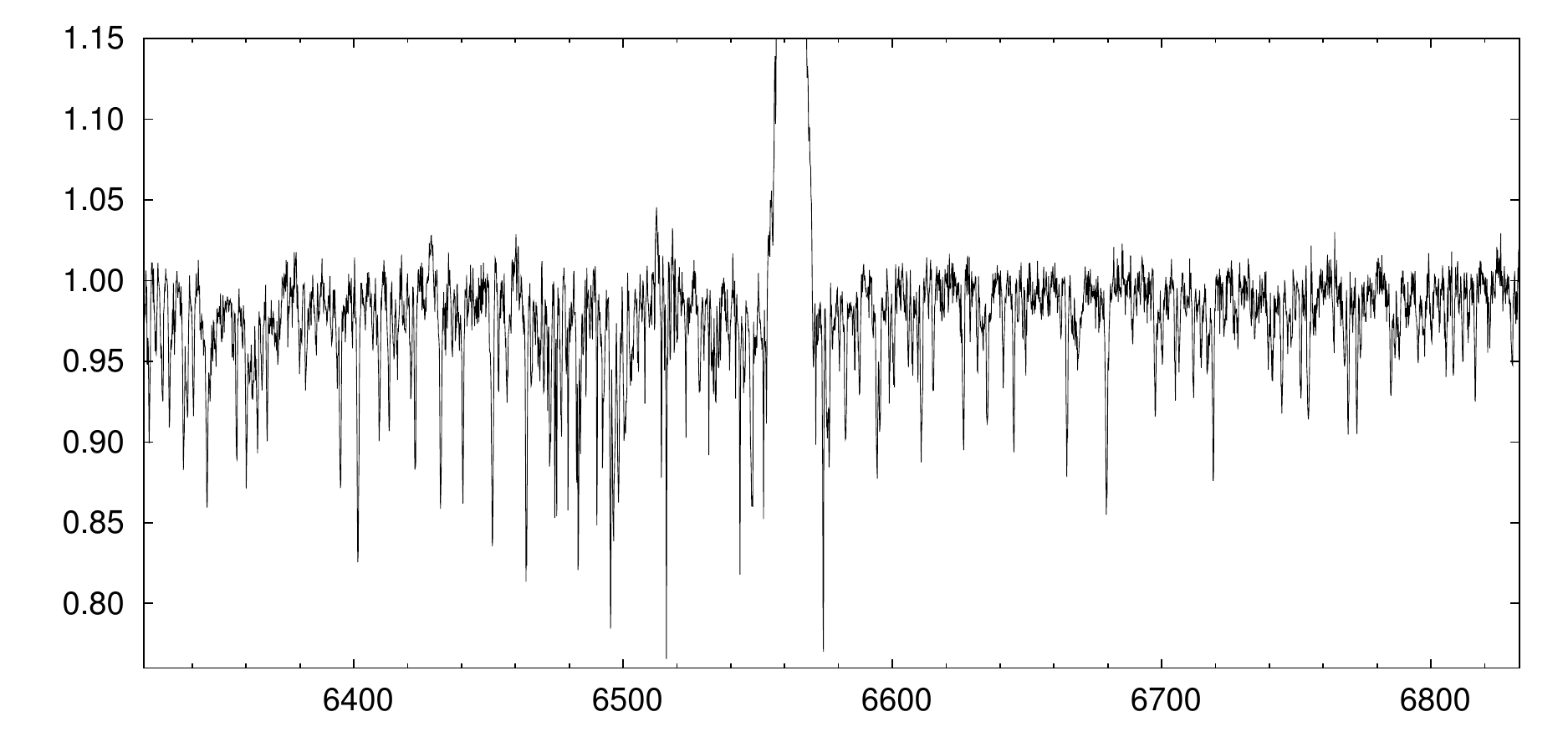}
\includegraphics[width=9.0cm]{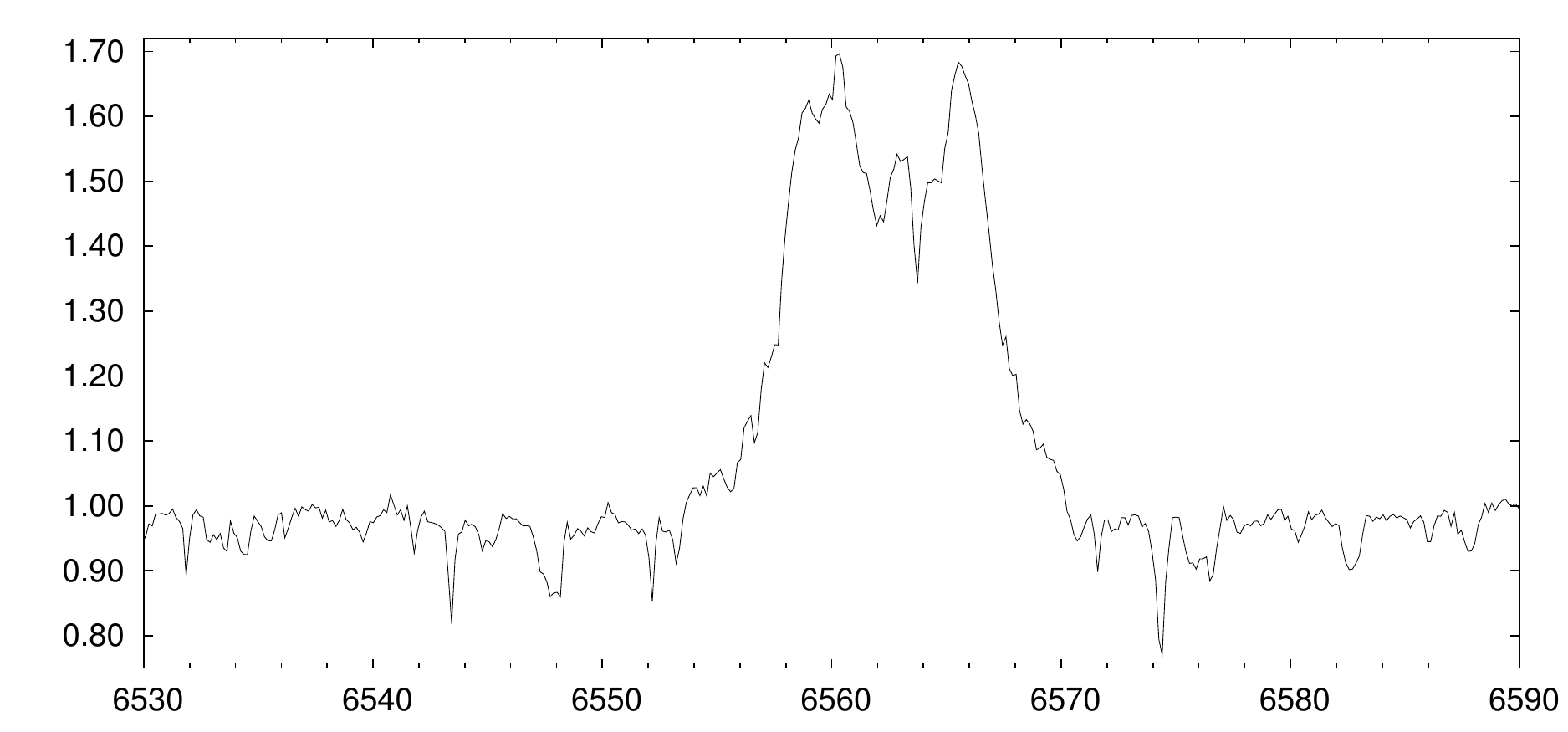}
\includegraphics[width=9.0cm]{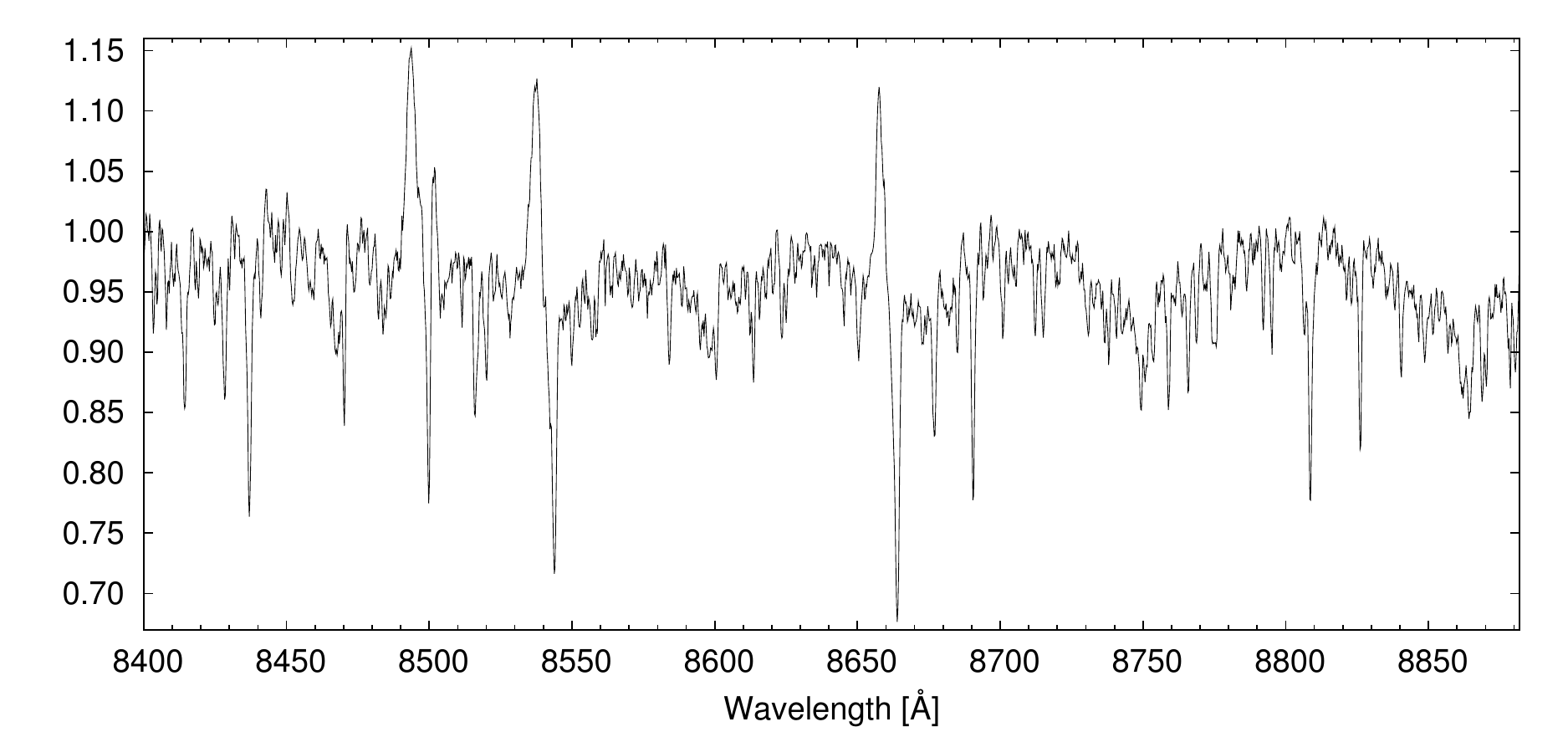}
\caption{Examples of available blue, red, and infrared spectra.
From top to bottom: blue spectrum, red spectrum, enlarged part of a red
spectrum near to \hae, and the infrared spectrum.
All three regions contain numerous lines of the cool component.}
\label{samples}
\end{figure}

Examples of the red, infrared and blue spectra at our disposal are in
Fig.~\ref{samples}. In all spectra numerous spectral lines typical for
a late spectral type are present. In addition to it,
one can see several \ion{H}{i} lines, affected by emission and a few absorption
lines belonging to the hotter component.

\subsection{Light and colour changes}
All light curves at our disposal exhibit double-wave ellipsoidal variations
with the orbital 33\fd8 period. Their amplitude is decreasing from the $R$
to the $U$ band, in which the changes are only barely visible. We show
the light curves later, along with the model curves.

\begin{figure}
\centering
\resizebox{\hsize}{!}{\includegraphics{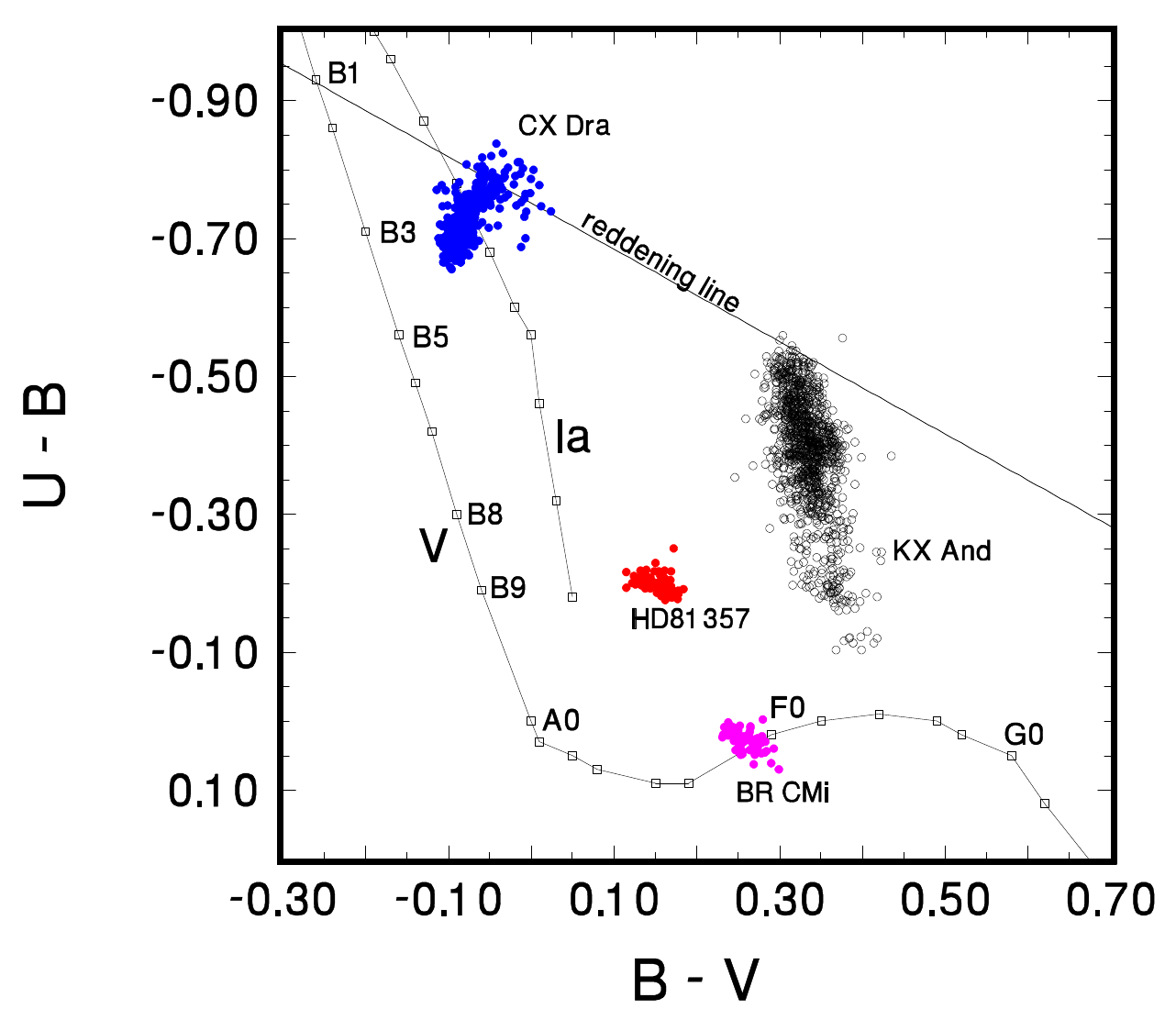}}
\caption{Variations of \hd in the colour - colour diagram are
compared to those known for some other Be stars observed at Hvar.}\label{ubv}
\end{figure}

  In Fig.~\ref{ubv} we compare the colour changes of \hd in the \ub\ vs.
\bv\ diagram with those known for several other well observed Be binary stars.
One can see that its colour changes are remarkably similar to those
for another ellipsoidal variable BR~CMi while other objects shown are
the representatives of the positive and inverse correlation between the light
and emission strength as defined by \citet{hvar83}
\citep[see also][]{bozic2013}. In particular, it is seen that for KX~And
dereddened colours exhibit inverse correlation with the object moving
along the main-sequence line in the colour-colour diagram from B1V to about
B7V. On the other hand, CX~Dra seems to exhibit a positive correlation
after dereddening, moving from B3\,V to B3\,I.

\subsection{Reliable radial velocities}
\subsubsection{\tt phdia RVs \label{sssec:spefo}}
Using the program {\tt phdia} (see Appendix~A), we measured
RVs of a number of unblended metallic lines in all blue, red and infrared spectra
at our disposal. A period analysis of these RVs confirmed and reinforced
the preliminary result of \citet{koubskytatry} that these RVs follow
a~sinusoidal RV curve with a~period of 33\fd77 and a semi-amplitude of 81~\ks.
As recently discussed in paper~30 \citep{zarfin30}, some caution must be
exercised when one analyses binaries with clear signatures
of the presence of circumstellar matter in the system. The experience
shows that the RV curve of the Roche-lobe filling component
is usually clean (with a possible presence of a Rossiter effect) and
defines its true orbit quite well, while many of the absorption
lines of the mass-gaining component are affected by the
presence of circumstellar matter, and their RV curves do not describe
the true orbital motion properly. It is therefore advisable to select
suitable spectral lines in the blue spectral region, free of such effects.

\subsubsection{\korel maps}
In the next step, we therefore derived a map of plausible solutions using
a~Python program kindly provided by J.A.~Nemravov\'a. The program employs
the program \korele, calculates true $\chi^2$ values and maps
the parameter space to find the optimal values of the semiamplitude $K_2$
and the mass ratio $M_1/M_2$. In our application to \hn, we omitted
the H$\gamma$ line and used the spectral region 4373 -- 4507~\AA,
which contains several \ion{He}{i} and metallic lines of the B-type
component~1. As in the case of a similar binary BR~CMi
\citep{zarfin30} we found that disentangling is not the optimal technique
how to derive the most accurate orbital solution since the interplay
between a~small RV amplitude of the broad-lined B primary and
its disentangled line  widths results in comparably good fits for
a~rather wide range of mass ratios. The lowest $\chi^2$ was obtained
for a mass ratio of 9.75, but there are two other shallow minima
near to the mass ratios 7 and 16. The optimal value of $K_2$ remained stable
near to 81 -- 82 \ks.

\subsubsection{Velocities derived via 2-D cross-correlation}
As mentioned earlier, the spectrum of the mass-gaining
component is usually affected by the presence of some contribution from
circumstellar matter, having a slightly lower temperature than the star
itself \citep[e.g.][]{desmet2010}. This must have impact on \korele,
which disentangles the composite spectra on the premise
that all observed spectroscopic variations arise solely from the orbital
motion of two binary components. Although we selected a blue spectral
region with the exclusion of the H$\gamma$ line (in the hope to minimize
the effect of circumstellar matter), it is probable that \korel solutions
returned spectra, which -- for the mass-gaining star -- average the true
stellar spectrum and a contribution from the circumstellar matter.

  As an alternative, we decided to derive RVs, which we anyway needed
to be able to combine photometry and spectroscopy in the \phoebe
program, with 2-D cross-correlation. We used the {\tt asTODCOR} program
written by one of us (YF) to this goal. The software is based on the method
outlined by \citet{todcor1} and has already been applied in similar cases
\citep[see][for the details]{desmet2010,zarfin30}. It performs
a 2-D cross-correlation between the composite observed spectra and template
spectra. The accuracy and precision of such measurements depend on both,
the quality of the observations, and on how suitable templates are chosen
to represent the contribution of the two stars.

\begin{figure}
\centering
\resizebox{\hsize}{!}{\includegraphics{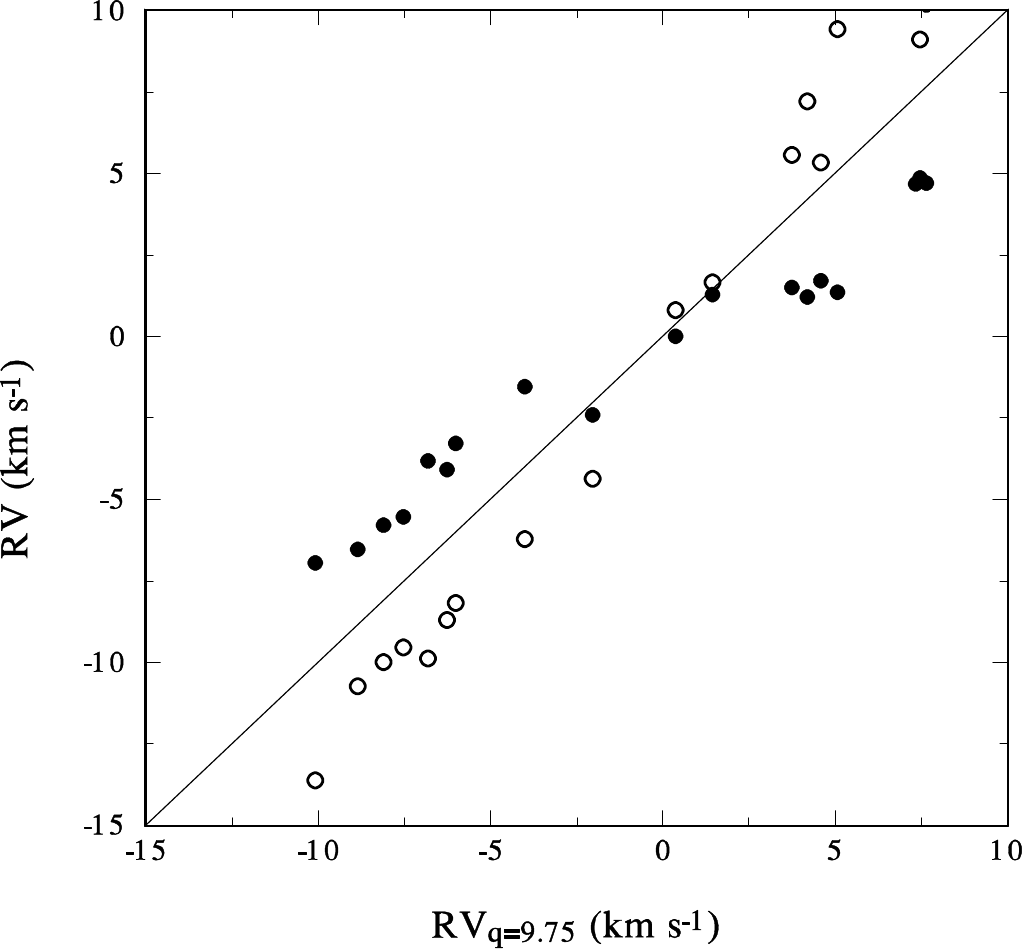}}
\caption{Comparison of {\tt asTODCOR} RVs of star~1 for the optimal
mass ratio q=9.75 (solid line) and for the two extreme mass-ratios \korel
templates (q=16 as filled circles, and q=7 as open circles).
Typical errors of {\tt asTODCOR} individual RVs are close to 1.0~\ks -- see
Table~\ref{rvtod}}
\label{minmax}
\end{figure}

In what follows, we used the observed blue spectra over the wavelength range from
4370 to 4503~\AA, mean resolution 0.12 \Ame, and a luminosity ratio
$L_2/L_1=0.1$. We adopted the spectra disentangled by \korel for
the optimal mass ratio 9.75 as the templates for the 2-D cross-correlation.
We attempted to investigate the effect of circumstellar matter on
the RVs derived with {\tt asTODCOR} for the primary using also the template
spectra for the other two mass ratios derived with \korele.
The RVs are compared in Fig.~\ref{minmax}.
From it we estimate that, depending on the
orbital phase, the systematic error due to the presence of circumstellar
matter may vary from 0 to 3 km/s.

The resulting {\tt asTODCOR} RVs for the optimal mass ratio 9.75 are listed,
with their corresponding random error bars, in Table~\ref{rvtod} in Appendix,
while the \spefo RVs of the cooler star are in Tab.~\ref{rvspefo}.

Assigning the three sets of {\tt asTODCOR} RVs from the blue
spectral region with the weights inversely proportional to the square of
their rms errors, we derived orbital solutions for them.
The solutions were derived separately for the hot component~1 and the cool
component~2 to verify whether those for the hot star describe its true orbital
motion properly. We used the program \fotel \citep{fotel1,fotel2}.
The results are summarised in
Table~\ref{trial}. We note that the values of the superior conjunction of
component~1 from the separate solutions for component~1 and component~2
agree within their estimated errors, but in all three cases the epochs
from component~1 precede a~bit those from component~2. This
might be another indirect indication that the RVs of component~1 are not
completely free of the effects of circumstellar matter.

A disappointing conclusion of this whole exercise is that there is no
reliable way how to derive a~unique mass ratio from the RVs. One has
to find out some additional constraints.

\subsection{Trial orbital solutions}

\begin{table}
\caption[]{Trial \fotel orbital solutions based on {\tt asTODCOR} RVs
from the 18 blue spectra, separately for star~1, and star~2.
Circular orbit was assumed and the period was kept fixed at 33\fd773~d.
The solutions were derived using three different RV sets based on the \korel
template spectra for the optimal mass ratio $q$ of 9.75, and for two other
plausible mass ratios 7, and 16. The epoch of the superior conjunction
$T_{\rm super.\, c.}$ is in RJD-55487 and rms is the rms error of
1 measurement of unit weight.}
\label{trial}
\begin{center}
\begin{tabular}{lcccccl}
\hline\hline\noalign{\smallskip}
Element                    &$q=9.75$     &$q=7$        &$q=16$ \\
\noalign{\smallskip}\hline\noalign{\smallskip}
  star 1 \\
\noalign{\smallskip}\hline\noalign{\smallskip}
$T_{\rm super.\, c.}$      &0.24(39)     &0.50(27)     &0.36(64)\\
$K_1$ (\ks)                &8.04(57)     &11.09(42)    &4.78(92)     \\
rms$_1$ (\ks)              &1.451        &1.505        &1.441        \\
\noalign{\smallskip}\hline\noalign{\smallskip}
  star 2 \\
\noalign{\smallskip}\hline\noalign{\smallskip}
$T_{\rm super.\, c.}$      &0.617(24)    &0.615(24)    &0.621(25)\\
$K_2$ (\ks)                &81.82(32)    &81.89(32)    &81.82(34)    \\
rms$_2$ (\ks)              &0.884        &0.887        &0.965        \\
\noalign{\smallskip}\hline\noalign{\smallskip}
\end{tabular}
\end{center}
\tablefoot{We note that due to the use of the \korel template spectra,
the {\tt asTODCOR} RVs are referred to zero systemic velocity.}
\end{table}

In the next step, we derived another orbital solution based on 151 RVs
(115 \spefo RVs, 18 {\tt asTODCOR} RVs of component~2 and 18 {\tt asTODCOR}
RVs of component~1).
As the spectra are quite crowded with numerous lines of component~2, we were
unable to use our usual practice of correcting the zero point of the velocity
scale via measurements of suitable telluric lines \citep{sef0}.
That is, why we allowed for the determination of individual systemic velocities
for the four subsets of spectra defined in Table~\ref{jourv}.
All RVs were used with the weights inversely proportional to the
square of their rms errors. This solution is in Table~\ref{fotelsol}.
There is a very good agreement in the systemic velocities from all
individual data subsets, even for the {\tt phdia} RVs from blue spectra,
where only four spectral lines could be measured and averaged.

\begin{table}
\caption[]{Trial \fotel orbital solutions based on all 151 {\tt phdia} and
{\tt asTODCOR} RVs and a solution for the 65 RVs of the \ha emission wings
measured in \spefoe. The epoch of superior conjunction of star~1 is
in RJD, rms is the rms error of 1 observation of unit weight.}
\label{fotelsol}
\begin{center}
\begin{tabular}{lcccccl}
\hline\hline\noalign{\smallskip}
Element           &  Binary & emis. wings \ha  \\
\noalign{\smallskip}\hline\noalign{\smallskip}
$P$ (d)                 &33.77458\p0.00079&33.77458 fixed \\
$T_{\rm super.\, c.}$   &55487.657\p0.020 &55492.53\p0.49 \\
$e$                     & 0.0 assumed     & 0.0 assumed   \\
$K_1$ (\ks)             &7.76\p0.56       & 9.47\p0.88    \\
$K_2$ (\ks)             &81.75\p0.17      & --            \\
$K_2/K_1$               &0.0949\p0.0063   & --            \\
$\gamma_1$ (\ks)        &$-$13.46\p0.18   &$-$11.33\p0.85 \\
$\gamma_2$ (\ks)        &$-$13.04\p0.24   & --            \\
$\gamma_3$ (\ks)        &$-$13.10\p1.38   & --            \\
$\gamma_4$ (\ks)        &$-$13.69\p0.25   &$-$13.13\p0.74 \\
$\gamma_{\rm 3T1}$ (\ks)&$-$0.39\p0.37$^*$&-- \\
$\gamma_{\rm 3T2}$ (\ks)&$-$0.20\p0.33$^*$&-- \\
rms        (\ks)     & 1.138 & 2.71           \\
\noalign{\smallskip}\hline\noalign{\smallskip}
\end{tabular}
\end{center}
\tablefoot{$^*)$ We note that the {\tt asTODCOR} RVs derived using the \korel templates refer to zero
systemic velocity.}
\end{table}

\subsection{Radiative properties of binary components}\label{synt}
To determine the radiative properties of the two binary components,
we used the Python program \pyte,
which interpolates in a pre-calculated grid of synthetic spectra. Using
a set of observed spectra, it tries to find the optimal fit between
the observed and interpolated model spectra with the help
of a~simplex minimization technique. It returns the radiative properties of
the system components such as \teff, \vsin or $\log~g$, but also the
relative luminosities of the stars and
RVs of individual spectra.

\begin{figure}
\centering
\resizebox{\hsize}{!}{\includegraphics{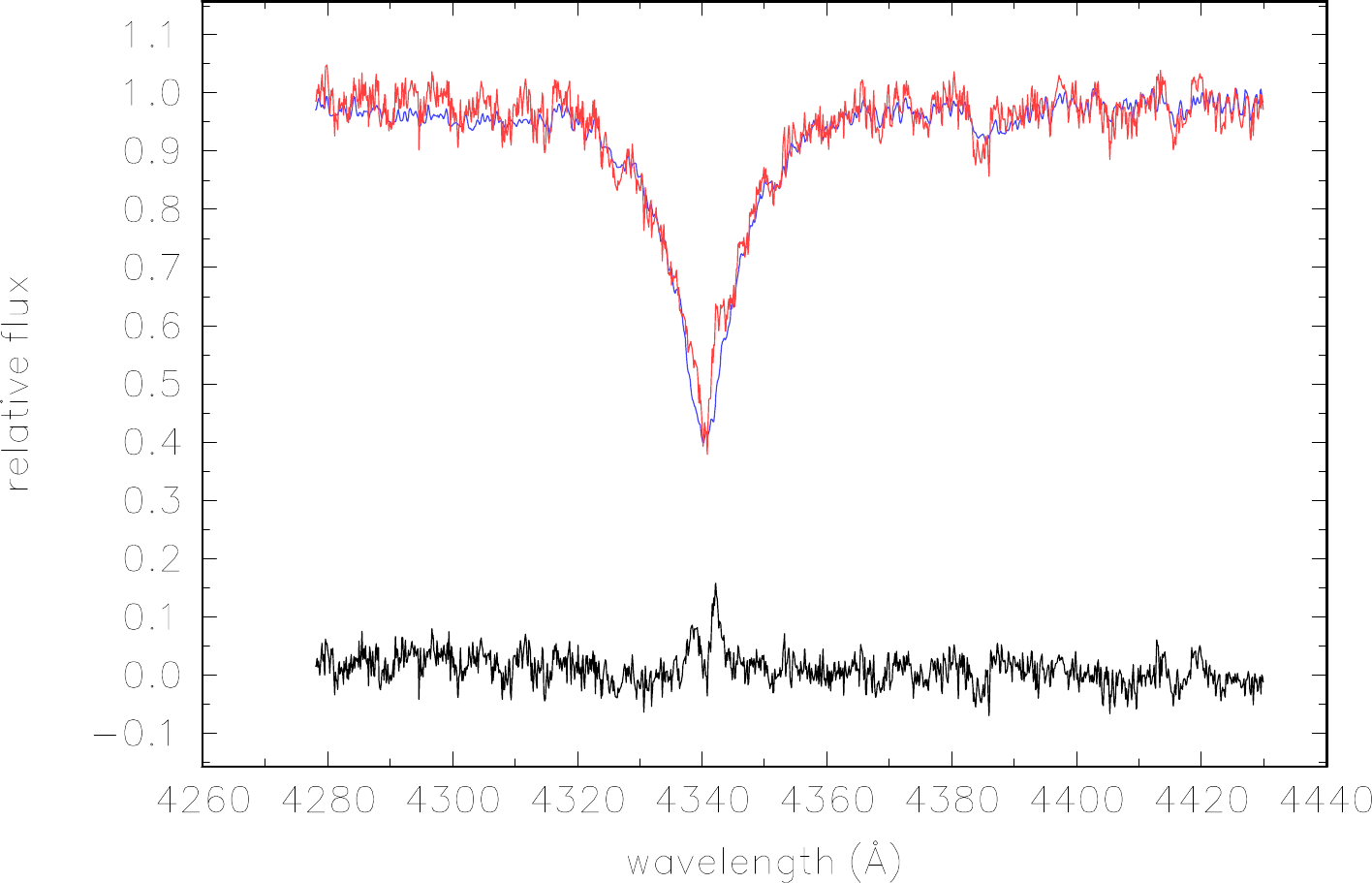}}
\resizebox{\hsize}{!}{\includegraphics{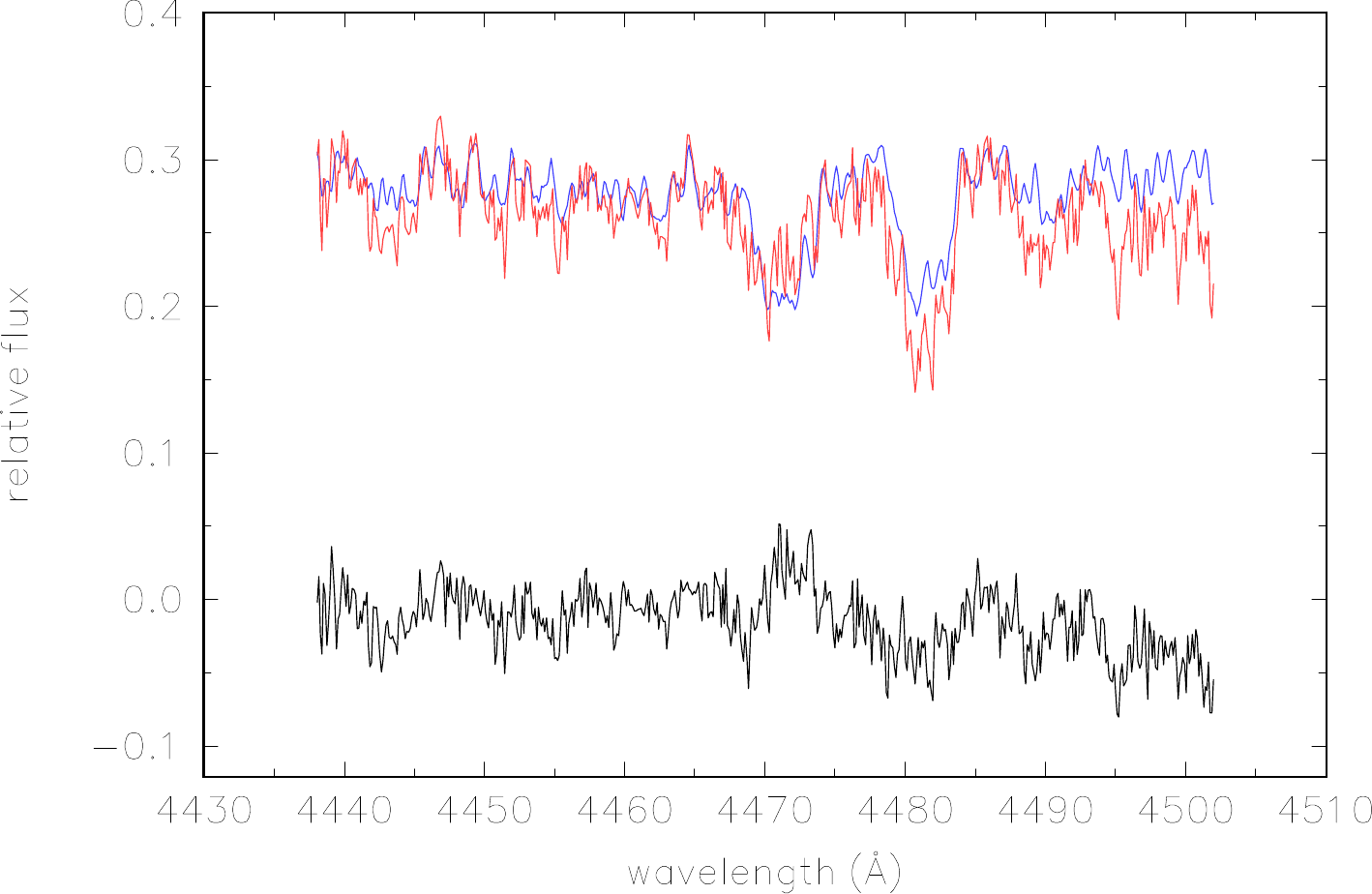}}
\caption{Example of the comparison of an~observed blue spectrum in
two selected spectral regions with a~combination of two synthetic spectra.
The residuals in the sense observed minus synthetic are also shown
on the same flux scale. To save space, the spectra in the bottom panel
were linearly shifted from 1.0 to 0.3\,.
The H$\gamma$ emission clearly stands out in the residuals in the first panel.
See the text for details.}
\label{synblue}
\end{figure}

\begin{table}
\caption[]{Radiative properties of both binary components derived from
a~comparison of selected wavelength segments of the observed and interpolated
synthetic spectra in the blue spectral region.}
\label{synpar}
\begin{center}
\begin{tabular}{rcccccl}
\hline\hline\noalign{\smallskip}
Element         & Component 1  & Component 2  \\
\noalign{\smallskip}\hline\noalign{\smallskip}
\tef (K)        &12930\p540    &4260\p24\\
\lgg [cgs]      &4.04\p0.13    &2.19\p0.14\\
$L_{4278-4430}$ &0.9570\p0.0092&0.0706\p0.0059\\
$L_{4438-4502}$ &0.9440\p0.0096&0.0704\p0.0075\\
\vsin (\ks)     &166.4\p1.5    &19.7\p0.8\\
\noalign{\smallskip}\hline\noalign{\smallskip}
\end{tabular}
\end{center}
\end{table}

In our particular application, two different grids of spectra were used:
AMBRE grid computed by \citet{delaverny2012} was used for component~2,
and the Pollux database computed by \citet{pala2010} was used for
component~1. We used the 18 file D blue spectra from OND,
which contain enough spectral lines of component~1.
The following two spectral segments, avoiding the region of the diffuse
interstellar band near to 4430~\AA, were modelled simultaneously:

\smallskip
\centerline{4278--4430~\AA, and 4438--4502~\AA.}

Uncertainties of radiative properties were obtained through
Markov chain Monte Carlo (MCMC) simulation implemented
within {\tt emcee}\footnote{The library is
available through GitHub~\url{https://github.com/dfm/emcee.git}
and its thorough description is
at~\url{http://dan.iel.fm/emcee/current/}.}
Python library by~\citet{fore2013}.
They are summarised in Table~\ref{synpar} and an example of the fit is in
Fig~\ref{synblue}.
We note that \pyt derives the RV from individual spectra without any
assumption about orbital motion. It thus represents some test of the
results obtained in the previous analysis based on {\tt asTODCOR} RVs.


\begin{table}
\centering
\caption{Combined radial-velocity curve and light curve solution
with \phoebee. The optimised parameters are given with only their
formal errors derived from the covariance matrix.
}
\label{phoebe}
\begin{tabular}{lrlrl}
\hline
Element& \multicolumn{4}{c}{Orbital properties}\\
\hline\hline
$P$\,(d)& \multicolumn{4}{c}{33.77445$\pm0.00041$}\\
$T_{\rm sup.c.}$\,(RJD)& \multicolumn{4}{c}{55487.647$\pm0.010$}\\
$M_1/M_2$\,& \multicolumn{4}{c}{10.0\p0.5}\\
$i$\,(deg)& \multicolumn{4}{c}{63$\pm5$}\\
$a$\,(\rs)& \multicolumn{4}{c}{63.01\p0.09}\\
\hline
\hline
& \multicolumn{4}{c}{Component properties}\\
& \multicolumn{2}{c}{Component 1}& \multicolumn{2}{c}{Component 2}\\
\hline
\tef\,(K)& \multicolumn{2}{c}{12930\p540}& \multicolumn{2}{c}{4260\p24}\\
$\log g_{\rm \left[cgs\right]}$& 3.75\p0.15& &1.71\p0.25\\
$M$\,(\Mnom)$^{*}$& 3.36\p0.15& & 0.34\p0.04\\
$R$\,(\Rnom)$^{*}$& 3.9\p0.2& &14.0\p0.7\\
$F$& 30.7\p1.9 & &1.0 fixed\\
$L_{\rm U}$      &$0.992\pm0.003$&&$0.008\pm0.003$\\
$L_{\rm B}$      &$0.956\pm0.003$&&$0.044\pm0.003$\\
$L_{\rm Hp}$     &$0.902\pm0.004$&&$0.098\pm0.004$\\
$L_{\rm V}$      &$0.853\pm0.005$&&$0.147\pm0.005$\\
$L_{\rm V(ASAS)}$&$0.853\pm0.005$&&$0.147\pm0.005$\\
$L_{\rm R}$      &$0.749\pm0.007$&&$0.251\pm0.007$\\
\hline
\end{tabular}
\tablefoot{
$^*$ Masses and radii are expressed in nominal solar units, see
\citet{prsa2016}.
}
\end{table}


\begin{table}
\caption{$\chi^2$ values for individual data sets for the adopted
combined \phoebe solution presented in Table~\ref{phoebe}.}\label{chi2}
\centering
\begin{tabular}{crccrrrlc}
\hline\hline
Data set       & No. of        &Original   & Normalised       \\
              & obs.          & $\chi^2$  &  $\chi^2$        \\
\hline
Hvar $U$      &   27          &   36.1    &     1.34         \\
Hvar $B$      &   27          &   33.0    &     1.22         \\
Hvar $V$      &   27          &   62.1    &     2.30         \\
ASAS-SN $V$   &  209          &  209.9    &     1.00         \\
Hipparcos $Hp$&  128          &  176.3    &     1.38         \\
Hvar $R$      &   15          &   22.7    &     1.51         \\
$RV_2$        &  116          &  304.1    &     2.62         \\
\hline
Total         &  549          & 1063.9    &     1.94         \\
\hline
\end{tabular}
\end{table}



\subsection{Combined light-curve and orbital solution in \phoebee}\label{psol}

To obtain the system properties and to derive the
final ephemeris, we used the program \phoebe~1 \citep{prsa2005,prsa2006}
and applied it to all photometric observations listed in Table~\ref{jouphot}
and {\tt phdia} and {\tt asTODCOR} RVs for star~2. Since \phoebe cannot
treat different systemic velocities, we used actually RVs minus respective
$\gamma$ velocities from the solution listed in Table~\ref{fotelsol}.
For the OND blue spectra (file~D of Table~\ref{jourv}), we omitted
the less accurate \spefo RVs and used only {\tt asTODCOR} RVs.
Bolometric albedos for star~1 and 2 were estimated from Fig.~7 of
\citet{claret2001} as 1.0 and 0.5, respectively. The coefficients of the
gravity darkening $\beta$ were similarly estimated as 1.0 and 0.6 from Fig.~7
of \citet{claret98}.

When we tried to model the light curves on the assumption that the secondary
is detached from the Roche lobe, we were unable to model the light-curve
amplitudes. We therefore conclude that \hd is a semi-detached binary
in the mass transfer phase between the components.

It is not possible to calculate the solution in the usual way. One
parameter, which comes into the game, is the synchronicity parameter $F$,
which is the ratio between the orbital and rotational period for each
component. While it is safe to adopt $F_2=1.0$ for the Roche-lobe filling
star~2, the synchonicity parameter $F_1$ must be re-caculated after
each iteration as

\begin{equation}
F_1=P_{\rm orbital}{v_1\sin i\over{50.59273R_1^{\rm e}\sin i}}\, \, ,
\end{equation}

\noindent where the equatorial radius $R_1^{\rm e}$ is again in the nominal
solar radius \Rnom, the orbital period in days, and the projected rotational
velocity in \ks. We adopt the value of 166.4~\kms for $v_1\sin i$ from the
\pyt solution.

It is usual that there is a very strong parameter degeneracy for
an ellipsoidal variable. To treat the problem we fixed \teff\ of both
components obtained from the \pyt solution and used
the very accurate parallax of \hd $p=0\farcs0016000 \pm 0\farcs0000345$
from the second data release of the Gaia satellite \citep{gaia2, dr2sum}
to restrict a~plausible range of the solutions.

From a~\phoebe light-curve solution for the Hvar \ubv\ photometry in
the standard Johnson system it was possible to estimate the following
\ubv\ magnitudes of the binary at light maxima

\smallskip
\centerline{$V_{1+2}=8$\m321, $B_{1+2}=8$\m491, and $U_{1+2}=8$\m300\, .}

\smallskip
\noindent We calculated a number of trial \phoebe solutions, keeping
the parameters of the solution fixed and mapping a~large parameter space.
For each such solution we used the resulting relative luminosities to
derive the \ubv\ magnitudes of the hot star~1, dereddened them in
a standard way and interpolating the bolometric correction from the
tabulation of \citet{flower96} we derived the range of possible values
for the mean radius $R_1$ for the Gaia parallax and its range of
uncertainty from the formula
\begin{equation}
M_{\rm bol\,1}=42.35326 - 5\log R_1 - 10\log T_{\rm eff\,1}
\end{equation}
\noindent \citep{prsa2016}. This range of the radius was compared to the
mean radius $R_1$ obtained from the corresponding \phoebe solution. We found
that the agreement between these two radius determinations could only be
achieved for a very limited range of mass ratios, actually quite close
to the mass ratio from the optimal \korel solution.

The resulting \phoebe solution is listed in Table~\ref{phoebe} and defines
the following linear ephemeris, which we shall adopt in the rest of this study
\begin{equation}
T_{\rm super.conj.}={\rm RJD}\,55487.647(10)+33\fd77445(41)\times\ E\,.
\label{efe}
\end{equation}

\begin{figure*}
\centering
\resizebox{\hsize}{!}{\includegraphics[angle=-90]{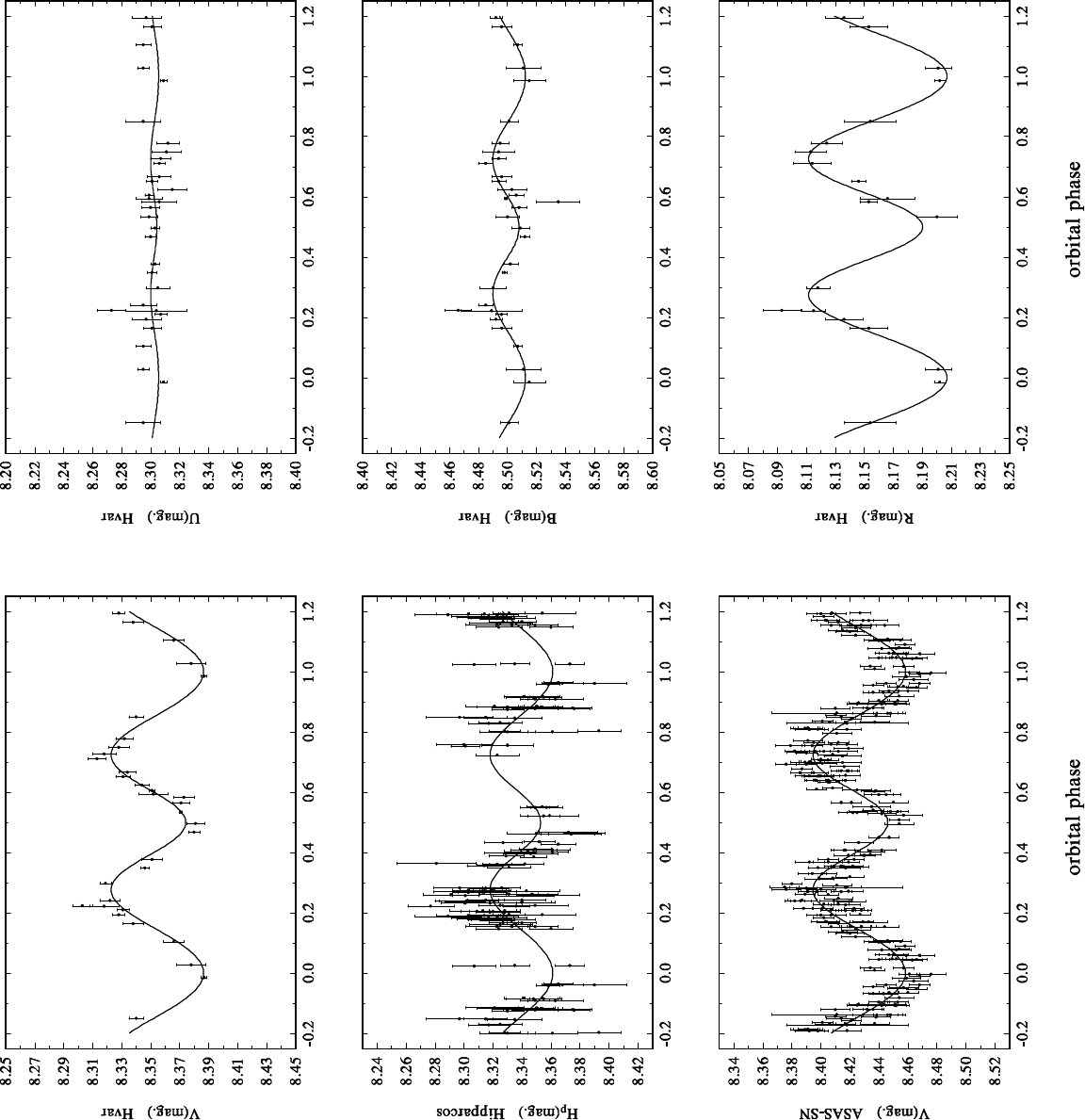}}
\caption{\ubvr\ light curves modelled with \phoebee.
Abscissa is labelled with orbital phases according to ephemeris
(\ref{efe}).The same scale on ordinate was used to show the changes
of the amplitude with passband. For all curves, also the rms errors of
individual data points are shown.}\label{lcphoebe}
\end{figure*}

\begin{figure}
\centering
\resizebox{\hsize}{!}{\includegraphics[angle=-90]{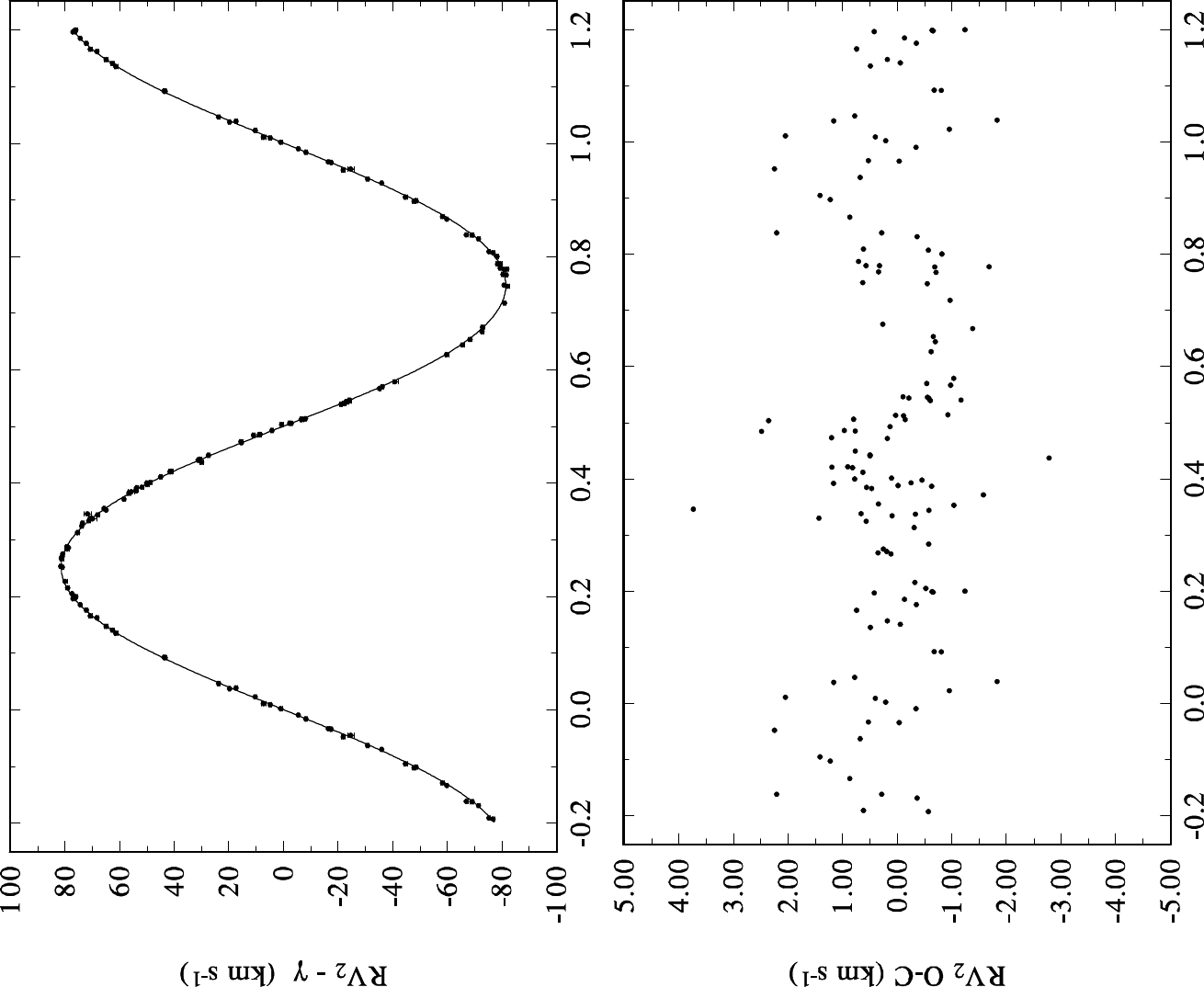}}
\resizebox{\hsize}{!}{\includegraphics[angle=-90]{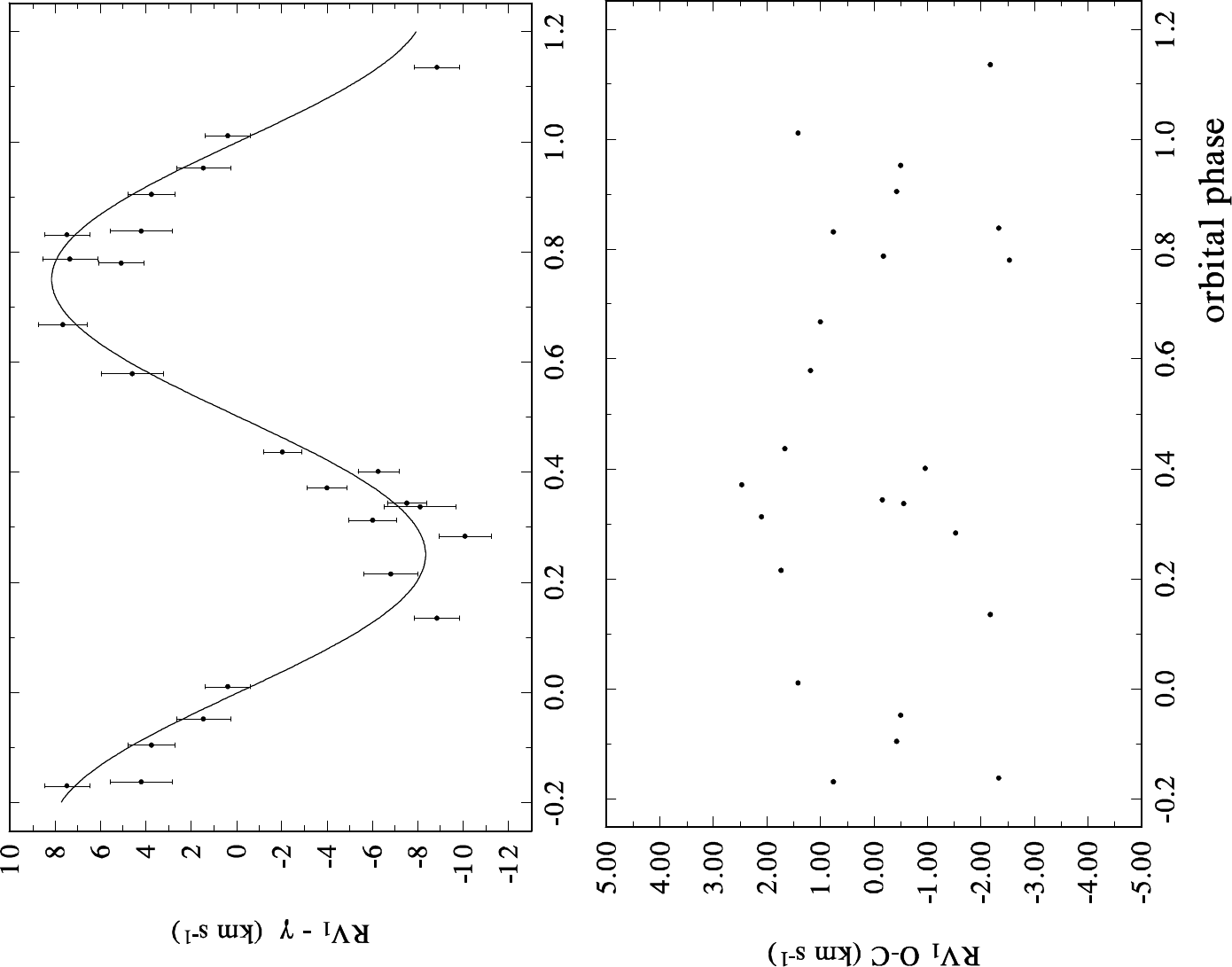}}
\caption{Top panels: RV curve of star~2 and the residuals from
\phoebe solution. The rms errors are comparable to the symbols' size.
Bottom panels: {\tt asTODCOR} RV curve of star~1, not used in the
\phoebe solution, and its residuals from that solution.}
\label{rvc}
\end{figure}

\noindent
The fit of the individual light curves is shown in Fig.~\ref{lcphoebe} while
in Fig.~\ref{rvc} we compare the fit of the RV curve of component 2 and also
the model RV curve of component~1 with the optimal {\tt asTODCOR} RVs, which
however  were not used in the \phoebe solution.

 The combined solutions of the light- and
RV-curves demonstrated a~strong degeneracy among individual
parameters. In Table~\ref{chi2} we show the original and normalized
$\chi^2$ values for the individual data sets used.  It is seen that
the contribution of the photometry and RVs to the total sum of squares
are comparable. A higher $\chi^2$ for RVs might be related to the fact
that we were unable to compensate small systematic differences in the
zero points of RVs between individual spectrographs and/or spectral
regions perfectly, having no control via telluric lines. The degeneracy
of the parameter space is illustrated by the fact that over a~large
range of inclinations and tolerable range of mass ratios the program
was always able to converge, with the total sum of squares differing
by less than three per cent.

In passing we note that we also independently tested the results from
\phoebe using the {\tt BINSYN} suite of programs
\citep{Linnell1984, Linnell1996, Linnell2000} with steepest descent method
to optimise the parameters of the binary system \citep{Sudar2011}.
This basically confirmed the results obtained with \phoebee.


\section{\ha profiles}

\begin{figure}
\centering
\resizebox{\hsize}{!}{\includegraphics{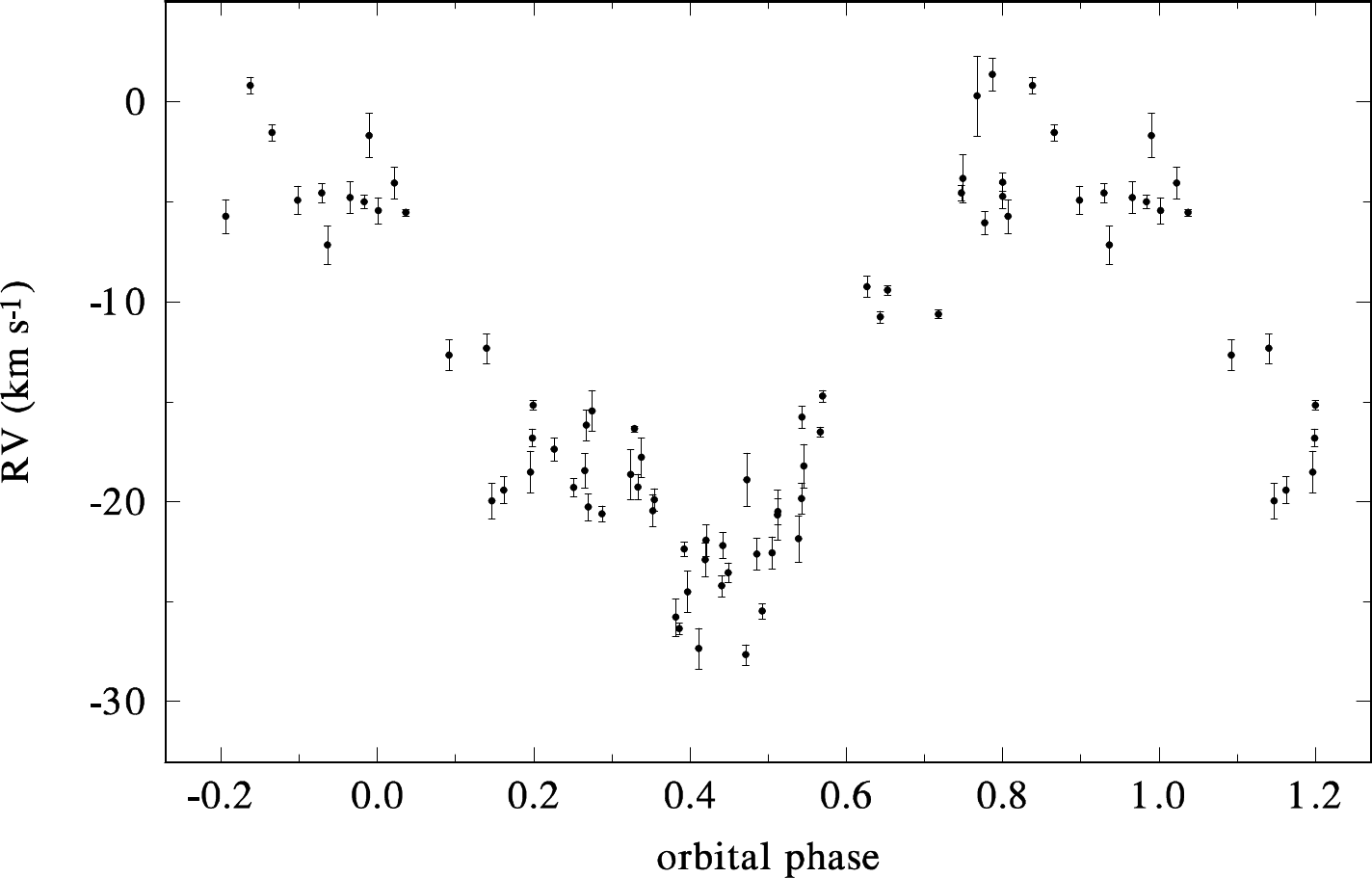}}
\caption{\spefo RVs of the \ha emission wings with their estimated
rms errors plotted vs. orbital phase.}
\label{RV_em_Ha}
\end{figure}

\begin{figure}
\centering
\resizebox{\hsize}{!}{\includegraphics{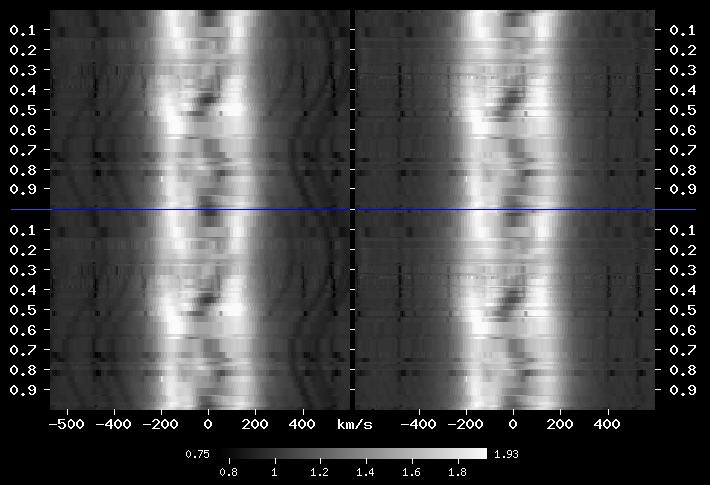}}
\caption{ Dynamical spectra in the neighbourhood of the \ha profile in
a grey-scale representation created in the program {\tt phdia}.
Left-hand panel: observed spectra;
right-hand panel: difference spectra after the subtraction
of a synthetic spectrum (including the \ha profile) of star~2.
The phases shown on both sides of the figure, correspond to ephemeris
(\ref{efe}). The spectra are repeated over two orbital cycles
to show the continuity of the orbital phase changes.}
\label{h3dyn}
\end{figure}

The strength of the \ha emission peaks ranges from 1.6 to 2.0 of the continuum
level and varies with the orbital phase (cf. Fig.~\ref{h3dyn}).
For all spectra a~central reversal in the emission is
present. But in many cases the absorption structure is quite complicated.

The RV curve of the \ha emission wings is well defined, sinusoidal and has
a~phase shift of some 5 days relative to the RV curve based on the lines of
component~1 (see Table~\ref{fotelsol} and Fig.~\ref{RV_em_Ha}).
This must be caused by some asymmetry in the distribution of the circumstellar
material producing the emission. In principle, however, one can conclude
that the bulk of the material responsible for \ha emission moves in orbit with
component~1 as it is seen from the semi-amplitude and systemic velocity
of the \ha emission RV curve, which are similar to those of the component~1.
We note that in both, the  magnitude and sense of the phase shift, this
behaviour is remarkably similar to that of another emission-line
semi-detached binary AU~Mon \citep{desmet2010}.

The orbital variation of the H$\alpha$ emission-line profiles
in the spectrum of HD~81357 is illustrated in the dynamical spectra
created with the program {\tt phdia} -- see the
left panel of Fig.~\ref{h3dyn}. The lines of the cool component are seen
both shorward and longward of the \ha emission. These lines can be used
to trace the motion of \ha absorption originating from star~2.
It is readily seen that the behaviour of the central absorption
in the \ha emission is more complicated than what would correspond
to the orbital motion of star~2. Three stronger absorption features
can be distinguished: region 1 near to phase 0.0, region 2 visible
from phase 0.4 to 0.5, and somewhat fainter feature 3 present
between phases 0.65 and 0.85. The absorption in region 2 follows
the motion of star~2, while the absortion line in region 3 moves
in antiphase.

In the right panel of Fig.~\ref{h3dyn} we show the dynamical spectra of
the difference profiles resulting after subtraction of an interpolated
synthetic spectrum of star~2
(properly shifted in RV according to the orbital motion
of star~2) from the observed \ha profiles. We note that the lines of
star~2 disappeared, but otherwise no pronounced changes in the \ha profiles
occurred in comparison to the original ones. Thus, the regions of
enhanced absorption 1, 2, and 3, already seen on the original profiles
are phase-locked and must be connected with the distribution of
circumstellar matter in the system.

 There are two principal possible geometries of the regions responsible
for the \ha emission: Either an accretion disk, the radius of which
would be limited by the dimension of the Roche lobe around the hot
star, $\sim 37$~\Rnom\ for our model;
or a bipolar jet perpendicular to the orbital plane, known, for instance, for
$\beta$~Lyr \citep{hec96}, V356~Sgr, TT~Hya, and RY Per \citep{peters2007}
or V393~Sco \citep{men2012b}, originating from the interaction of the
gas stream encircling the hot star with the denser stream flowing from
the Roche-lobe filling star~2 towards star~1 \citep{bis2000}.
In passing we note that no secular change in the strength of the \ha
emission over the interval covered by the data could be detected.

\section{Stellar evolution of \hd with mass exchange}
Given a rather low mass and a large radius of the secondary,
we were interested to test whether stellar evolution with mass exchange
in a binary can produce a~system similar to \hn.
We use the binary stellar evolution program MESA \citep{pax2011,pax2015}
in order to test a certain range of input parameters.
We tried the initial masses in the intervals
$M_1 \in (1.0; 1.5)\,\Mnom$,
$M_2 \in (2.2; 2.7)\,\Mnom$,
and the initial binary period
$P \in (2; 10)\,{\rm days}$.
Hereinafter, we use the same notation as in the preceding text,
so that $M_1$ is the original secondary, which becomes the primary during
the process of mass exchange.
The mass transfer was computed with
\cite{Ritter_1988A&A...202...93R} explicit scheme,
with the rate limited to $\dot M_{\rm max} = 10^{-7}\,\Mnom,{\rm yr}^{-1}$,
and magnetic breaking of \cite{Rappaport_1983ApJ...275..713R},
with the exponent $\gamma = 3$.
For simplicity, we assumed zero eccentricity, conservative mass transfer,
no tidal interactions, and no irradiation.
We used the standard time step controls.

\begin{figure}
\centering
\includegraphics[width=9.0cm]{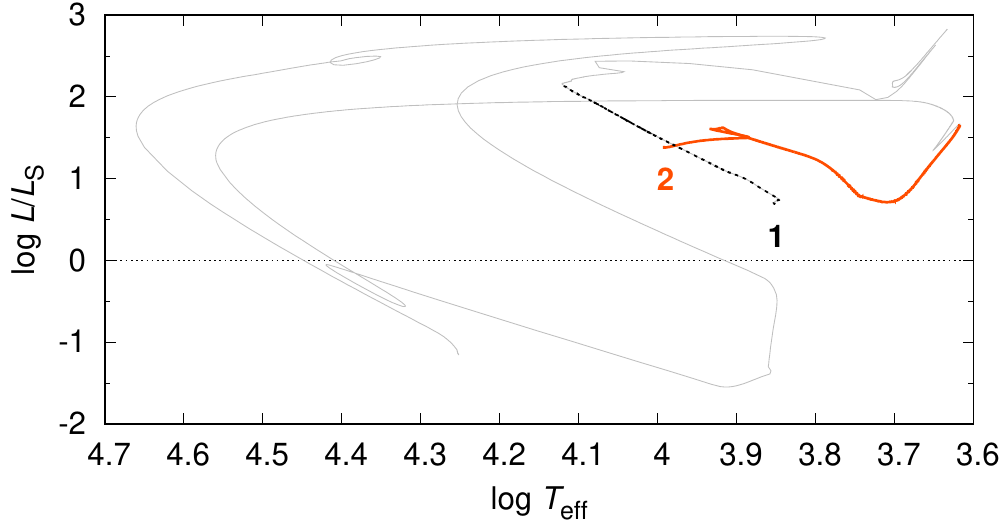}
\includegraphics[width=9.0cm]{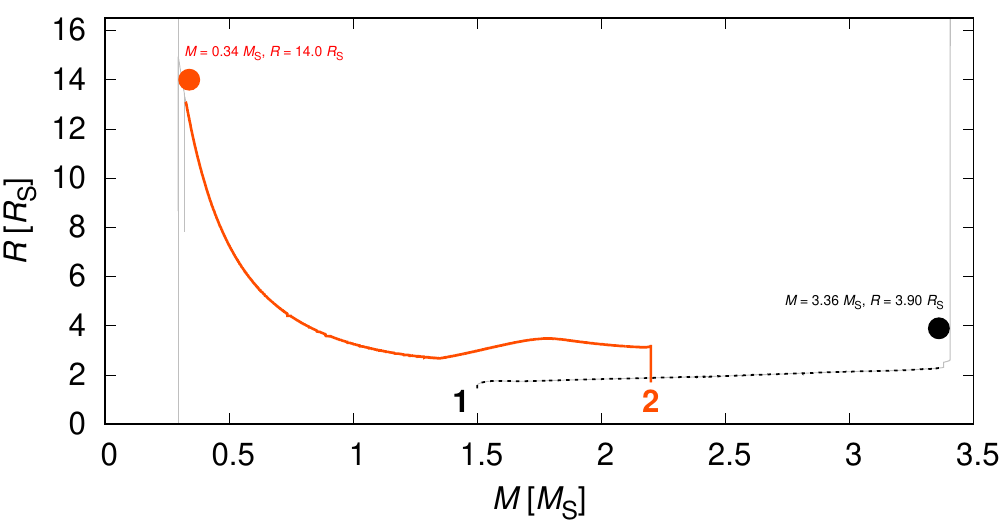}
\includegraphics[width=9.0cm]{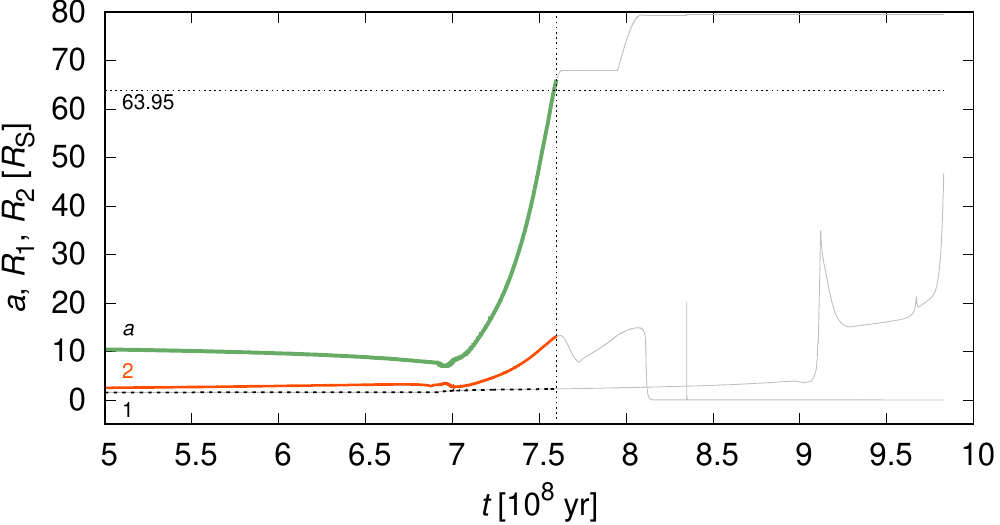}
\caption{Long-term evolution of HD~81357 binary
as computed by MESA \citep{pax2015}.
The initial masses were
$M_1 = 1.5\,M_\odot$,
$M_2 = 2.2\,M_\odot$,
and the initial period
$P = 2.4\,{\rm d}$.
Top: the HR diagram with the
(resulting) primary denoted as 1 (dashed black line),
secondary as 2 (solid orange line)
at the beginning of the evolutionary tracks (ZAMS).
Middle: the radius~$R$ vs mass~$M$;
the corresponding observed values are also indicated (filled circles).
Bottom: the semimajor axis~$a$ of the binary orbit (green)
and the radii~$R_1$, $R_2$ of the components
vs time~$t$.
The observed value $a = 63.95\,R_\odot$
and the time $t \doteq 7.6\cdot10^8\,{\rm yr}$ with the maximal~$R_2$
are indicated (thin dotted).
Later evolutionary phases are also plotted (solid grey line).
}
\label{HRD}
\end{figure}

An example for the initial masses
$M_1 = 1.5\,\Mnom$,
$M_2 = 2.2\,\Mnom$,
the initial period
$P = 2.4\,{\rm d}$
and the mass transfer beginning on the SGB
is presented in Figure~\ref{HRD}.
We obtained a result, which matches the observations surprisingly well,
namely the final semimajor axis
$a_{\rm syn} = 66.04\,\Rnom$,
which corresponds to the period
$P_{\rm syn} \doteq 32.60\,{\rm d}$ (while $P_{\rm obs} = 33.77\,{\rm d}$),
the final masses
$M_1 = 3.37\,\Mnom$ ($3.36\,\Mnom$),
$M_2 = 0.33\,\Mnom$ ($0.34\,\Mnom$),
the maximum secondary radius
$R_2 = 13.1\,\Rnom$ ($14.0\,\Rnom$),
with the exception of the primary radius
$R_1 = 2.3\,\Rnom$ (cf. $3.9\,\Rnom$).
 Alternatively, solutions can be also found for different ratios
of the initial masses $M_1$, $M_2$,
and later phases of mass transfer (RGB),
although they are sometimes preceded by an overflow.
An advantage may be an even better fit of the final~$R_1$,
and a relatively longer duration of the inflated~$R_2$
which makes such systems more likely to be observed.

Consequently, we interpret the secondary as a low-mass star with a still
inflated envelope, close to the end of the mass transfer phase.
We demonstrated that a binary with an ongoing mass transfer
is a reasonable explanation for both components of the \hd system.
A more detailed modelling including an accretion disk,
as carried out by \citet{VanRensbergen_2016A&A...592A.151V}
for other Algols, would be desirable.

\section{Discussion}

 Quite many of similar systems with a~hot mass-gaining component were found
to exhibit cyclic long-term light and colour variations, with cycles
an order of magnitude longer than the respective orbital periods
\citep[cf. a~good recent review by][]{mennic2017}, while others seem
to have a~constant brightness outside the eclipses. The later seems to be
the case of \hn, for which we have not found any secular changes.
It is not clear as yet what is the principal factor causing the
presence of the cyclic secular changes
\citep[see also the discussion in][]{zarfin30}.
A few yellow-band photoelectric observations of \hd found in the literature
all give 8\m3 to 8\m4. It is notable, however, that both magnitudes
given in the HD catalogue are 7\m9. Future monitoring of the system
brightness thus seems desirable.

  We note that we have not found any indication of a secular period change
and also our modelling of the system evolution and its low mass ratio
$M_2/M_1$ seem to indicate that \hd is close to the end of mass exchange
process. The same is also true about BR~CMi studied by \citet{zarfin30}.
We can thus conjecture that the cyclic light variations mentioned above
do not occur close to the end of the mass exchange.

\begin{acknowledgements}
We gratefully aknowledge the use of the latest publicly available versions
of the programs \fotel and \korel written by P.~Hadrava. Our sincere
thanks are also due to A.Pr\v{s}a, who provided us with a modified version of
the program \phoebe 1.0 and frequent consultations on its usage, and to
the late A.P.~Linnell for his program suite BINSYN.
We also thank J.A.~Nemravov\'a, who left the astronomical
research in the meantime, for her contributions to this study and for the
permission to use her Python programs, \pyt and several auxiliary ones,
and to C.S.~Kochanek for the advice on the proper use of the ASAS-SN
photometry. Suggestions and critical remarks by an anonymous referee
helped us to re-think the paper, improve and extend the analyses, and
clarify some formulations.
A.~Aret, \v{S}.~Dvo\v{r}\'akov\'a, R.~K\v{r}i\v{c}ek, J.A.~Nemravov\'a,
P.~Rutsch, K.~\v{S}ejnov\'a, and P.~Zasche obtained a few Ond\v{r}ejov
spectra, which we use. The research of PK was supported from
the ESA PECS grant 98058. The research of PH and MB was supported by
the grants GA15-02112S and GA19-01995S of the Czech Science Foundation.
HB, DR, and DS acknowledge financial support from
the Croatian Science Foundation under the project 6212 ``Solar and
Stellar Variability", while the research of D.~Kor\v{c}\'akov\'a is supported
from the grant GA17-00871S of the Czech Science Foundation.
This work has made use of data from the European Space Agency (ESA)
mission {\it Gaia} (\url{https://www.cosmos.esa.int/gaia}), processed by
the {\it Gaia} Data Processing and Analysis Consortium (DPAC,
\url{https://www.cosmos.esa.int/web/gaia/dpac/consortium}). Funding
for the DPAC has been provided by national institutions, in particular
the institutions participating in the {\it Gaia} Multilateral Agreement.
We gratefully acknowledge the use of the electronic databases: SIMBAD at CDS,
Strasbourg and NASA/ADS, USA.
\end{acknowledgements}

\bibliographystyle{aa}
\bibliography{citace}

\Online
\begin{appendix}
\section{Details of the spectral data reduction and measurements}\label{apa}
The initial reduction of all OND and DAO spectra (bias subtraction,
flat-fielding, creation of 1-D spectra, and wavelength calibration) was
carried out in {\tt IRAF}. Optimal extraction was used.
Normalization, removal of residual cosmics and flaws and RV measurements of
\ha emission profiles were carried out with the program \spefo
\citep{sef0,spefo}, namely the latest version 2.63 developed
by Mr.~J.~Krpata \citep{spefo3}. \spefo displays direct and flipped traces of
the line profiles superimposed on the computer screen that the user can slide
to achieve a precise overlapping of the parts of the profile of whose RV
one wants to measure. RVs of a selection of stronger unblended lines of
the cool component~2 (see Table~\ref{klist}) covering the red and infrared
spectral regions (available for all spectra), were measured using the program
{\tt phdia} written by LK. The program {\tt phdia} is a~web application which
allows the creation of dynamical spectra and, using a similar principle as
the program \spefoe, also RV measurements based on the comparison of direct
and flipped line profiles. The input is a~normalized digital spectrum.
A robust mean RV with the rms error was derived for each spectrum to obtain
the mean RV. In contrast to our usual approach,
we have not measured a~selection of good telluric lines to obtain
an additional fine correction of the RV zero point of each spectrogram
because of a strong blending of the telluric lines with the stellar lines of
the cool component~2. Moreover, we measured the RVs of the steep wings of the
\ha emission line repeatably at different times to obtain an estimate
of the rms error for these measurements. We point out
that although some broad and shallow lines of component~1 are seen
in the spectra, their direct RV measurement is impossible because of
numerous blends with the lines of component~2.

\begin{table}
\caption[]{List of spectral lines of component~2 and their air
wavelengths used for the RV measurements in {\tt phdia}. The numbering of
the multiplets of individual ions corresponds to the one introduced by
\citet{moore45}. The wavelengths are taken from the NIST Atomic Spectra
Database. In some cases, weighted mean of the wavelengths of close
blends in the neighborhood of the line in question was used
(such cases are denoted by letter "c").}
\label{klist}
\begin{center}
\begin{tabular}{clclcccl}
\hline\hline\noalign{\smallskip}
Wavelength   & Element & Wavelength   & Element \\
(\AA)       &         &  (\AA)      \\
\noalign{\smallskip}\hline\noalign{\smallskip}
              \\
4425.824&\ion{Ti}{i}  78      &6624.912&\ion{Fe}{i}   13 c\\
4427.310&\ion{Fe}{i}   2      &6633.749&\ion{Fe}{i} 1197  \\
4430.189&\ion{Fe}{i} 472      &6643.638&\ion{Ni}{i}   43  \\
4454.780&\ion{Ca}{i}   4      &6663.442&\ion{Fe}{i}  111  \\
6322.685&\ion{Fe}{i} 207      &6677.986&\ion{Fe}{i}  268 c\\
6327.450&\ion{Cr}{i}          &6717.690&\ion{Ca}{i} 1194  \\
6338.999&\ion{Fe}{i} 1258 c   &8426.504&\ion{Ti}{i}   33  \\
6344.149&\ion{Fe}{i}  169     &8468.407&\ion{Fe}{i}   60  \\
6355.029&\ion{Fe}{i}  342     &8582.257&\ion{Fe}{i}  401  \\
6358.697&\ion{Fe}{i}   13     &8611.803&\ion{Fe}{i}  339  \\
6366.424&\ion{Ti}{i}  103 c   &8621.600&\ion{Fe}{i}  401  \\
6393.601&\ion{Fe}{i}  168     &8648.462&\ion{Si}{i}       \\
6400.147&\ion{Fe}{i}  816+13 c&8662.140&\ion{Ca}{ii}      \\
6408.018&\ion{Fe}{i}  816     &8674.985&\ion{Fe}{i}       \\
6411.649&\ion{Fe}{i}  816     &8688.625&\ion{Fe}{i}   60  \\
6419.872&\ion{Fe}{i} 1258     &8710.391&\ion{Fe}{i} 1267  \\
6421.350&\ion{Fe}{i}  111     &8757.187&\ion{Fe}{i}  339  \\
6430.846&\ion{Fe}{i}   62     &8763.966&\ion{Fe}{i} 1172  \\
6439.070&\ion{Ca}{i}   18     &8793.343&\ion{Fe}{i} 1172  \\
6462.658&\ion{Ca}{i}   18     &8806.757&\ion{Mg}{i}    7  \\
6471.660&\ion{Ca}{i}   18     &8824.221&\ion{Fe}{i}   60  \\
6613.756&\ion{Fe}{i}    c     &8838.428&\ion{Fe}{i}  339  \\
        &                     &8866.932&\ion{Fe}{i} 1172  \\
\noalign{\smallskip}\hline\noalign{\smallskip}
\end{tabular}
\end{center}
\end{table}

\begin{table*}
\caption[]{Individual {\tt asTODCOR} RVs of the hot component~1 and
cool component~2 derived from the 18 file~D spectra in the blue spectral
region 4370 - 4503~\AA\ (columns 2 and 3), using the \korel template
for $q=0.975$, and {\tt phdia} RVs of component~2 (column 4).}
\label{rvtod}
\begin{center}
\begin{tabular}{crrrcccccccl}
\hline\hline\noalign{\smallskip}
RJD        &  RV$_1$ \ \ \ \ & RV$_2$ \ \ \ \ & RV$_2^{\spefoe}$\\
           &(\ks)\ \ \ \  &(\ks)\ \ \ \  & (\ks)\\
\noalign{\smallskip}\hline\noalign{\smallskip}
56765.3585  &$  7.47\pm0.99$ &$-71.72\pm0.61$  &$-94.00\pm5.15$\\
56769.4557  &$  1.45\pm1.19$ &$-22.17\pm0.69$  &$-30.83\pm3.91$\\
56771.4295  &$  0.38\pm0.99$ &$  7.07\pm0.80$  &$ -3.78\pm2.68$\\
56778.3392  &$ -6.81\pm1.19$ &$ 78.73\pm0.63$  &$ 70.07\pm5.14$\\
56782.4491  &$ -8.11\pm1.59$ &$ 69.74\pm1.59$  &   --          \\
56797.4027  &$  5.08\pm1.01$ &$-79.91\pm0.48$  &$-96.32\pm1.82$\\
56799.3612  &$  4.20\pm1.36$ &$-69.34\pm0.80$  &$-85.33\pm0.65$\\
56809.4039  &$ -8.85\pm0.99$ &$ 61.11\pm0.68$  &$ 48.81\pm0.97$\\
56814.4188  &$-10.09\pm1.14$ &$ 78.99\pm0.73$  &$ 63.61\pm1.66$\\
56815.4057  &$ -6.01\pm1.05$ &$ 75.10\pm0.52$  &$ 61.20\pm1.94$\\
56816.4472  &$ -7.53\pm0.88$ &$ 67.75\pm0.72$  &$ 59.85\pm2.34$\\
56817.3872  &$ -4.00\pm0.88$ &$ 58.16\pm0.55$  &$ 45.85\pm3.74$\\
56818.3936  &$ -6.26\pm0.90$ &$ 48.52\pm0.60$  &$ 40.45\pm1.35$\\
56824.3882  &$  4.59\pm1.36$ &$-41.11\pm1.04$  &$-49.37\pm1.46$\\
56827.3853  &$  7.66\pm1.08$ &$-73.04\pm0.60$  &$-85.43\pm4.14$\\
56831.4217  &$  7.35\pm1.20$ &$-78.75\pm0.83$  &$-88.40\pm1.99$\\
56835.3919  &$  3.75\pm1.05$ &$-44.89\pm0.68$  &$-55.12\pm4.26$\\
56853.3661  &$ -2.04\pm0.85$ &$ 29.70\pm0.68$  &$ 25.21\pm6.86$\\
\noalign{\smallskip}\hline\noalign{\smallskip}
\end{tabular}
\end{center}
\tablefoot{We remind that the {\tt asTODCOR} RVs are referred to zero
systemic velocity.}
\end{table*}

\begin{table*}
\caption[]{Individual RVs of the  cool component~2 and of the \ha emission
wings (RV$_{\rm H\alpha\ em.}$) from the red and IR spectra measured in
{\tt phdia} and \spefoe.
Individual files are identified by their file codes from Table~\ref{jourv}.}
\label{rvspefo}
\begin{center}
\begin{tabular}{crrccrrcccccl}
\hline\hline\noalign{\smallskip}
RJD&RV$_2$ \ \ &RV$_{\rm H\alpha\ em.}$& Spg.&RJD&RV$_2$\ \ &RV$_{\rm H\alpha\ em.}$&Spg.\\
          &\ \ (\ks)   & \ \  (\ks)   & &           &\ \ (\ks)&\ \ (\ks) &    \\
\noalign{\smallskip}\hline\noalign{\smallskip}
55621.5844&$-31.00\pm0.31$&$ -4.79\pm0.77$&A/1&55740.4180&$ -2.17\pm0.60$&  --   &C/2\\
55644.4775&$-79.07\pm0.63$&$-10.76\pm0.29$&A/1&55750.3954&$-92.50\pm0.63$&  --   &C/2\\
55661.4801&$ 51.43\pm0.56$&$-19.97\pm0.87$&A/1&55751.3787&$-88.36\pm0.40$&  --   &C/2\\
55672.4335&$  1.89\pm0.41$&$-27.67\pm0.50$&A/1&55754.3639&$-60.93\pm0.71$&  --   &C/2\\
55691.5362&$  6.19\pm0.35$&$ -5.54\pm0.17$&A/1&55759.3896&$ 10.68\pm0.50$&  --   &C/2\\
55693.4009&$ 30.16\pm0.46$&$-12.68\pm0.77$&A/1&55808.6067&$-12.40\pm0.62$&  --   &C/2\\
55703.3408&$ 40.36\pm0.45$&$-26.35\pm0.28$&A/1&55817.5587&$-93.63\pm0.86$&  --   &C/2\\
55707.3493&$-16.37\pm0.66$&$-22.57\pm0.81$&A/1&55831.6385&$ 61.31\pm0.46$&  --   &C/2\\
55742.4201&$-36.56\pm0.56$&$-15.79\pm0.56$&A/1&56007.4786&$ 40.66\pm0.50$&  --   &C/2\\
55749.3717&$-94.36\pm0.52$&$ -3.83\pm1.21$&A/1&56012.4986&$-35.32\pm0.63$&  --   &C/2\\
55752.3664&$-80.57\pm0.56$&$  0.82\pm0.41$&A/1&56044.4130&$ -4.28\pm0.57$&  --   &C/2\\
55807.5902&$  2.00\pm0.63$&$-18.92\pm1.33$&A/1&56045.3781&$-21.21\pm0.43$&  --   &C/2\\
55817.5303&$-95.16\pm0.51$&$  0.28\pm2.01$&A/1&56046.4416&$-37.25\pm0.65$&  --   &C/2\\
55836.5069&$ 60.10\pm0.48$&$-16.36\pm0.15$&A/1&56642.6600&$ 63.16\pm0.32$&  --   &C/2\\
55877.5068&$-36.76\pm0.56$&$-19.85\pm0.78$&A/1&56771.3714&$ -8.29\pm0.49$&  --   &C/2\\
55969.5233&$ 67.82\pm0.60$&$-16.18\pm0.76$&A/1&56797.3361&$-94.94\pm0.56$&  --   &C/2\\
56007.5227&$ 38.30\pm0.45$&$-22.38\pm0.34$&A/1&56810.4372&$ 57.67\pm0.55$&  --   &C/2\\
56008.4721&$ 27.50\pm0.46$&$-21.95\pm0.77$&A/1&56818.3462&$ 36.93\pm0.58$&  --   &C/2\\
56011.5609&$-20.15\pm0.66$&$-20.67\pm1.24$&A/1&57070.6612&$-71.38\pm0.60$&  --   &C/2\\
56013.4972&$-49.70\pm0.48$&$-14.73\pm0.28$&A/1&57073.4943&$-37.73\pm1.04$&  --   &C/2\\
56015.4132&$-73.31\pm0.70$&$ -9.26\pm0.54$&A/1&57117.3562&$ 68.45\pm0.46$&  --   &C/2\\
56044.4414&$ -4.95\pm0.58$&$-22.62\pm0.79$&A/1&57118.4756&$ 65.64\pm0.38$&  --   &C/2\\
56045.3434&$-20.13\pm0.56$&$-20.51\pm0.65$&A/1&55680.7562&$-94.71\pm0.48$&$-10.63\pm0.21$&E/4\\
56046.4681&$-37.62\pm0.52$&$-18.22\pm1.08$&A/1&55681.7524&$-95.77\pm0.60$&$ -4.56\pm0.40$&E/4\\
56235.4193&$ 49.07\pm0.65$&$-12.34\pm0.74$&A/1&55682.7582&$-94.65\pm0.50$&$ -6.07\pm0.58$&E/4\\
56433.3847&$-12.56\pm0.54$&$ -5.46\pm0.65$&A/1&55685.7612&$-73.64\pm0.44$&$ -1.54\pm0.40$&E/4\\
56642.6053&$ 63.54\pm0.43$&$-18.53\pm1.04$&B/1&55735.7331&$ 58.02\pm1.40$&    --         &E/4\\
56712.5061&$ 67.66\pm0.51$&$-18.45\pm0.84$&B/1&55859.9161&$ -3.46\pm0.47$&$ -4.07\pm0.79$&E/4\\
56718.4552&$ 17.03\pm0.36$&$-22.20\pm0.66$&B/1&55953.9582&$-90.56\pm0.45$&$ -5.74\pm0.85$&E/4\\
56746.4230&$ 67.52\pm0.50$&$-20.29\pm0.66$&B/1&55993.9278&$-19.21\pm0.35$&$ -1.69\pm1.11$&E/4\\
56759.3533&$-81.82\pm0.71$&$ -9.43\pm0.21$&B/1&56034.7448&$ 62.65\pm0.34$&$-16.83\pm0.44$&E/4\\
56764.3167&$-91.72\pm0.55$&$ -4.03\pm0.46$&B/1&56034.7806&$ 62.22\pm0.48$&$-15.19\pm0.24$&E/4\\
57074.4891&$-21.74\pm0.61$&$ -5.00\pm0.34$&B/1&56072.7588&$ 60.14\pm0.51$&$-18.65\pm1.24$&E/4\\
57080.5153&$ 54.84\pm0.50$&$-19.44\pm0.67$&B/1&56073.7163&$ 51.22\pm0.40$&$-20.46\pm0.82$&E/4\\
57101.6119&$-92.96\pm0.57$&$  1.36\pm0.82$&B/1&56073.7926&$ 51.93\pm0.40$&$-19.91\pm0.56$&E/4\\
57105.3867&$-62.06\pm0.51$&$ -4.92\pm0.70$&B/1&56074.7968&$ 42.03\pm1.12$&    --         &E/4\\
57106.4352&$-49.53\pm0.39$&$ -4.57\pm0.45$&B/1&56113.7894&$-34.91\pm0.44$&$-21.87\pm1.18$&E/4\\
57116.4673&$ 66.46\pm0.64$&$-17.40\pm0.58$&B/1&56296.0908&$-44.65\pm0.44$&$ -7.17\pm0.96$&E/4\\
57117.3111&$ 67.46\pm0.52$&$-19.30\pm0.46$&B/1&56379.9514&$ 27.72\pm0.51$&$-22.91\pm0.86$&E/4\\
57118.5416&$ 65.79\pm0.52$&$-20.62\pm0.40$&B/1&56380.9409&$ 13.71\pm0.37$&$-23.57\pm0.49$&E/4\\
55621.6176&$-29.52\pm0.40$&  --           &C/2&56384.9173&$-48.98\pm0.39$&$-16.53\pm0.25$&E/4\\
55645.5398&$-86.02\pm0.64$&  --           &C/2&56408.8349&$ 67.06\pm0.50$&$-15.47\pm1.02$&E/4\\
55662.4553&$ 59.10\pm0.56$&  --           &C/2&56410.8205&$ 57.49\pm0.49$&$-19.28\pm0.64$&E/4\\
55663.4181&$ 64.29\pm0.43$&  --           &C/2&56444.7391&$ 56.99\pm0.58$&$-17.79\pm0.98$&E/4\\
55691.5822&$  4.30\pm0.64$&  --           &C/2&56446.7468&$ 36.02\pm0.61$&$-24.52\pm1.04$&E/4\\
55693.3831&$ 30.23\pm0.50$&  --           &C/2&56787.7054&$ -9.58\pm0.50$&$-25.48\pm0.37$&E/4\\
55703.3784&$ 40.97\pm0.62$&  --           &C/2&56817.7530&$ 42.84\pm0.52$&$-25.79\pm0.92$&E/4\\
55704.4937&$ 28.33\pm0.63$&  --           &C/2&56818.7358&$ 31.25\pm0.48$&$-27.35\pm1.02$&E/4\\
55707.3759&$-15.42\pm0.57$&  --           &C/2&56819.7342&$ 17.45\pm0.52$&$-24.22\pm0.53$&E/4\\
\noalign{\smallskip}\hline\noalign{\smallskip}
\end{tabular}
\end{center}
\end{table*}

\section{Details on the photometric data used}\label{apb}
Below, we provide some details of the individual data sets and their reductions.
\begin{itemize}
\item Station 01 -- Hvar: \ \
 These differential \ubv\ and later \ubvr\ observations have been secured
by HB, PH, and DR relative to HD~82861 (the check star HD~81772 being
observed as frequently as the variable)
and carefully transformed to the standard $UBV(R)$ system via non-linear
transformation formul\ae\ using the {\tt HEC22} reduction program -- see
\citet{hhj94} and \citet{hechor98} for the observational strategy and
data reduction. \footnote{The whole program suite with a detailed manual,
examples of data, auxiliary data files, and results is available at
{\sl http://astro.troja.mff.cuni.cz/ftp/hec/PHOT}\,.}
All observations were reduced with the latest
{\tt HEC22 rel.18} program, which allows the time variation of
linear extinction coefficients to be modelled in the course of observing
nights. For the light-curve solutions we used normal points averaged
over the typical observing sequence of 0\fd055 and the corresponding
rms errors.
\item Station 61 -- Hipparcos: \ \ These all-sky observations were
reduced to the standard $V$ magnitude via the transformation formul\ae\
derived by \citet{hpvb} to verify that no secular light changes in
the system were observed. However, for the light-curve solution
in \phoebee, we consider the Hipparcos transmission curve for the $H_p$
magnitude and also use the original rms errors.
\item Station 93 -- ASAS-SN: \ \ These all-sky automated survey
for supernovae $V$ observations were adopted from the data server
https://asan-sn.osu.edu \citep{asas2014,asas2017} and cleaned for some
strongly deviating data points. We used only the observations
from the bb camera, which were numerous enough. The rms errors provided by
the on the fly calculator are unrealistically small \citep[cf.][]{jay2019}
and we used them only to assign relative weights to individual observations,
inversely proportional to their square and estimated the mean rms scatter
as 0.0065 mag.
\end{itemize}

Journal of all data sets is in Table~\ref{jouphot}.

\begin{table*}
\caption[]{Individual \ubvr\ observations from the Hvar Observatory.
The comparison star was HD 82861, for which we used the mean Hvar all-sky
values $V=7$\m073, \bv\ $=$ 0\m141, \ub\ $=$ 0\m093, and \vr\ $=$ 0\m129. $X$ is
the air mass of the observation and \tria X = X$_{\rm var.}-$X$_{\rm comp.}$
is the air-mass difference between the variable and comparison star.}
\label{ubvr}
\begin{center}
\begin{tabular}{cccccccccrc}
\hline\hline\noalign{\smallskip}
RJD       &weight&$V$&$B$&$U$&$R$&\bv&\ub&X&\tria X \ \ \\
          &    &(mag.)  &(mag.)&(mag.) &(mag.)&(mag.)&(mag.) \\
\noalign{\smallskip}\hline\noalign{\smallskip}
55879.6237&1.50& 8.349& 8.509& 8.301&  --  & 0.160&-0.208& 1.088&-0.005\\
55879.6307&1.50& 8.351& 8.507& 8.296&  --  & 0.156&-0.211& 1.079&-0.004\\
55879.6340&1.50& 8.353& 8.501& 8.300&  --  & 0.148&-0.201& 1.075&-0.004\\
55881.6675&1.50& 8.339& 8.499& 8.311&  --  & 0.160&-0.188& 1.042& 0.002\\
55881.6724&1.50& 8.334& 8.489& 8.299&  --  & 0.155&-0.190& 1.040& 0.002\\
55881.6756&1.50& 8.330& 8.499& 8.308&  --  & 0.169&-0.191& 1.039& 0.003\\
55938.4882&1.50& 8.348& 8.500& 8.304&  --  & 0.152&-0.196& 1.061&-0.002\\
55938.4946&1.50& 8.347& 8.497& 8.299&  --  & 0.150&-0.198& 1.056&-0.001\\
55938.4994&1.50& 8.343& 8.497& 8.301&  --  & 0.154&-0.196& 1.052&-0.000\\
55939.4333&1.50& 8.351& 8.497& 8.302&  --  & 0.146&-0.195& 1.137&-0.011\\
55939.4403&1.50& 8.345& 8.503& 8.301&  --  & 0.158&-0.202& 1.124&-0.009\\
55939.4438&1.50& 8.356& 8.505& 8.306&  --  & 0.149&-0.199& 1.118&-0.009\\
55942.4879&1.50& 8.383& 8.515& 8.297&  --  & 0.132&-0.218& 1.052&-0.000\\
55942.4944&1.50& 8.377& 8.511& 8.300&  --  & 0.134&-0.211& 1.048& 0.000\\
55942.4976&1.50& 8.380& 8.510& 8.304&  --  & 0.130&-0.206& 1.046& 0.001\\
55943.4734&1.50& 8.378& 8.510& 8.301&  --  & 0.132&-0.209& 1.063&-0.002\\
55943.4797&1.50& 8.387& 8.513& 8.306&  --  & 0.126&-0.207& 1.057&-0.001\\
55943.4829&1.50& 8.379& 8.504& 8.303&  --  & 0.125&-0.201& 1.054&-0.001\\
56001.3532&1.50& 8.332& 8.492& 8.307&  --  & 0.160&-0.185& 1.037& 0.003\\
56001.3600&1.50& 8.333& 8.498& 8.303&  --  & 0.165&-0.195& 1.036& 0.004\\
56001.3633&1.50& 8.327& 8.497& 8.310&  --  & 0.170&-0.187& 1.035& 0.004\\
56002.3781&1.50& 8.323& 8.481& 8.294&  --  & 0.158&-0.187& 1.035& 0.006\\
56002.3842&1.50& 8.327& 8.489& 8.303&  --  & 0.162&-0.186& 1.037& 0.007\\
56002.3873&1.50& 8.315& 8.484& 8.289&  --  & 0.169&-0.195& 1.038& 0.007\\
56013.3318&0.50& 8.379& 8.503& 8.293&  --  & 0.124&-0.210& 1.035& 0.004\\
56013.3364&0.50& 8.361& 8.505& 8.297&  --  & 0.144&-0.208& 1.035& 0.005\\
56013.3394&0.50& 8.372& 8.515& 8.310&  --  & 0.143&-0.205& 1.035& 0.005\\
56015.3404&0.50& 8.337& 8.499& 8.320&  --  & 0.162&-0.179& 1.035& 0.006\\
56015.3451&0.50& 8.349& 8.518& 8.327&  --  & 0.169&-0.191& 1.036& 0.006\\
56015.3482&0.50& 8.347& 8.491& 8.299&  --  & 0.144&-0.192& 1.037& 0.007\\
56065.3231&1.00& 8.359& 8.508& 8.291&  --  & 0.149&-0.217& 1.186& 0.017\\
56065.3294&1.00& 8.366& 8.504& 8.293&  --  & 0.138&-0.211& 1.202& 0.018\\
56065.3324&1.00& 8.373& 8.510& 8.300&  --  & 0.137&-0.210& 1.211& 0.018\\
56086.3506&1.00& 8.314& 8.498& 8.307&  --  & 0.184&-0.191& 1.516& 0.024\\
56086.3555&1.00& 8.327& 8.496& 8.314&  --  & 0.169&-0.182& 1.543& 0.024\\
56086.3590&1.00& 8.312& 8.488& 8.301&  --  & 0.176&-0.187& 1.563& 0.025\\
56747.3537&1.00& 8.313& 8.481& 8.295& 8.147& 0.168&-0.186& 1.040& 0.008\\
56747.3712&1.00& 8.319& 8.489& 8.308& 8.133& 0.170&-0.181& 1.050& 0.009\\
56747.3754&1.00& 8.322& 8.498& 8.310& 8.148& 0.176&-0.188& 1.053& 0.010\\
56755.3233&1.00& 8.370& 8.508& 8.303& 8.228& 0.138&-0.205& 1.037& 0.007\\
56755.3318&1.00& 8.371& 8.497& 8.299& 8.232& 0.126&-0.198& 1.040& 0.008\\
56755.3359&1.00& 8.370& 8.492& 8.292& 8.207& 0.122&-0.200& 1.042& 0.008\\
56757.3444&1.00& 8.340& 8.497& 8.288& 8.169& 0.157&-0.209& 1.051& 0.009\\
56757.3523&1.00& 8.357& 8.499& 8.305& 8.194& 0.142&-0.194& 1.057& 0.010\\
56757.3563&1.00& 8.357& 8.499& 8.300& 8.206& 0.142&-0.199& 1.061& 0.011\\
56759.2984&0.50& 8.332& 8.500& 8.306& 8.177& 0.168&-0.194& 1.035& 0.005\\
56759.3074&0.50& 8.336& 8.489& 8.298& 8.170& 0.153&-0.191& 1.036& 0.006\\
56759.3130&0.50& 8.323& 8.486& 8.296& 8.163& 0.163&-0.190& 1.037& 0.007\\
56761.3048&1.00& 8.310& 8.480& 8.301& 8.149& 0.170&-0.179& 1.037& 0.007\\
56761.3107&1.00& 8.308& 8.485& 8.308& 8.124& 0.177&-0.177& 1.038& 0.007\\
\noalign{\smallskip}\hline\noalign{\smallskip}
\end{tabular}
\end{center}
\end{table*}

\setcounter{table}{0}
\begin{table*}
\caption {continued}
\begin{center}
\begin{tabular}{cccccccccrc}
\hline\hline\noalign{\smallskip}
RJD       &weight&$V$&$B$&$U$&$R$&\bv&\ub&X&\tria X \ \ \\
          &    &(mag.)  &(mag.)&(mag.) &(mag.)&(mag.)&(mag.) \\
\noalign{\smallskip}\hline\noalign{\smallskip}
56761.3175&1.00& 8.320& 8.489& 8.306& 8.141& 0.169&-0.183& 1.041& 0.008\\
56804.3396&1.00& 8.387& 8.502& 8.309& 8.221& 0.115&-0.193& 1.309& 0.021\\
56804.3462&1.00& 8.384& 8.515& 8.310& 8.229& 0.131&-0.205& 1.334& 0.022\\
56804.3533&1.00& 8.386& 8.525& 8.306& 8.225& 0.139&-0.219& 1.362& 0.022\\
56812.3428&1.00& 8.319& 8.497& 8.313& 8.135& 0.178&-0.184& 1.414& 0.023\\
56812.3445&1.00& 8.327& 8.505& 8.316& 8.148& 0.178&-0.189& 1.422& 0.023\\
56812.3506&1.00& 8.307& 8.464& 8.279& 8.132& 0.157&-0.185& 1.451& 0.023\\
56858.3315&1.00& 8.377& 8.538& 8.320& 8.173& 0.161&-0.218& 2.307& 0.014\\
56858.3393&1.00& 8.364& 8.518& 8.301& 8.182& 0.154&-0.217& 2.403& 0.010\\
56858.3441&1.00& 8.375& 8.547& 8.297& 8.170& 0.172&-0.250& 2.464& 0.007\\
56867.3311&1.00& 8.334& 8.503& 8.286& 8.161& 0.169&-0.217& 2.622&-0.002\\
56867.3372&1.00& 8.344& 8.506& 8.310& 8.168& 0.162&-0.196& 2.709&-0.007\\
56867.3393&1.00& 8.339& 8.493& 8.293& 8.195& 0.154&-0.200& 2.740&-0.009\\
56873.3170&0.50& 8.390& 8.505& 8.289& 8.224& 0.115&-0.216& 2.652&-0.003\\
56873.3190&0.50& 8.363& 8.496& 8.299& 8.231& 0.133&-0.197& 2.681&-0.005\\
56873.3257&0.50& 8.379& 8.529& 8.300& 8.206& 0.150&-0.229& 2.782&-0.012\\
57100.3264&1.00& 8.325& 8.492& 8.311& 8.145& 0.167&-0.181& 1.045& 0.001\\
57100.3281&1.00& 8.316& 8.483& 8.303& 8.135& 0.167&-0.180& 1.044& 0.001\\
57100.3327&1.00& 8.327& 8.486& 8.300& 8.125& 0.159&-0.186& 1.042& 0.002\\
57100.3344&1.00& 8.328& 8.485& 8.304& 8.124& 0.157&-0.181& 1.041& 0.002\\
57100.3409&1.00& 8.334& 8.500& 8.320& 8.153& 0.166&-0.180& 1.038& 0.003\\
57100.3426&1.00& 8.333& 8.511& 8.324& 8.137& 0.178&-0.187& 1.038& 0.003\\
57101.2961&1.00& 8.326& 8.488& 8.307& 8.153& 0.162&-0.181& 1.068&-0.003\\
57101.2978&1.00& 8.321& 8.488& 8.303& 8.136& 0.167&-0.185& 1.067&-0.003\\
57101.3064&1.00& 8.332& 8.495& 8.313& 8.154& 0.163&-0.182& 1.058&-0.001\\
57101.3082&1.00& 8.339& 8.501& 8.326& 8.160& 0.162&-0.175& 1.057&-0.001\\
57101.3157&1.00& 8.334& 8.493& 8.307& 8.146& 0.159&-0.186& 1.051&-0.000\\
57101.3210&1.00& 8.335& 8.502& 8.314& 8.134& 0.167&-0.188& 1.047& 0.001\\
57114.3361&1.00& 8.335& 8.487& 8.301& 8.187& 0.152&-0.186& 1.036& 0.006\\
57114.3454&1.00& 8.330& 8.493& 8.293& 8.179& 0.163&-0.200& 1.039& 0.007\\
57114.3521&1.00& 8.351& 8.489& 8.297& 8.189& 0.138&-0.192& 1.041& 0.008\\
57114.3588&1.00& 8.340& 8.503& 8.299& 8.155& 0.163&-0.204& 1.045& 0.009\\
57114.3662&1.00& 8.336& 8.504& 8.300& 8.168& 0.168&-0.204& 1.050& 0.009\\
57114.3730&1.00& 8.335& 8.496& 8.310& 8.183& 0.161&-0.186& 1.055& 0.010\\
57115.3537&1.00& 8.332& 8.493& 8.286& 8.156& 0.161&-0.207& 1.044& 0.008\\
57115.3605&1.00& 8.330& 8.498& 8.310& 8.170& 0.168&-0.188& 1.048& 0.009\\
57115.3669&1.00& 8.348& 8.490& 8.294& 8.156& 0.142&-0.196& 1.052& 0.010\\
57115.3734&1.00& 8.334& 8.487& 8.287& 8.165& 0.153&-0.200& 1.058& 0.010\\
57115.3799&1.00& 8.331& 8.491& 8.297& 8.174& 0.160&-0.194& 1.064& 0.011\\
57116.3552&0.50& 8.294& 8.454& 8.264& 8.205& 0.160&-0.190& 1.046& 0.009\\
57116.3614&0.50& 8.313& 8.481& 8.289& 8.220& 0.168&-0.192& 1.050& 0.009\\
57116.3676&0.50& 8.294& 8.455& 8.257& 8.246& 0.161&-0.198& 1.055& 0.010\\
57116.3755&0.50& 8.308& 8.472& 8.277& 8.237& 0.164&-0.195& 1.062& 0.011\\
\noalign{\smallskip}\hline\noalign{\smallskip}
\end{tabular}
\end{center}
\end{table*}

\end{appendix}
\end{document}